\documentclass[twocolumn,aps,superscriptaddress]{revtex4-1}
\usepackage{palatino}
\usepackage[latin1]{inputenc}
\usepackage{epsf}
\usepackage{amsmath,amssymb}
\usepackage{latexsym}
\usepackage{calc}
\usepackage{color}

\usepackage{graphicx}
\usepackage{hyperref}
\newcommand{\ben}{\begin{equation}}
\newcommand{\een}{\end{equation}}
\newcommand{\bea}{\begin{eqnarray}}
\newcommand{\eea}{\end{eqnarray}}

\def\sss{\scriptscriptstyle\rm}

\def\1s{_{1,\sss S}}
\def\2s{_{2,\sss S}}
\def\x{_{\sss X}}
\def\c{_{\sss C}}
\def\s{_{\sss S}}
\def\xc{_{\sss XC}}

\def\Hxc{_{\sss HXC}}

\def\ext{_{\rm ext}}

\def\br{{\bf r}}

\begin{document}
 \title{Curing the Divergence in Time-Dependent Density Functional Quadratic Response Theory}
 \author{Davood Dar}
 \affiliation{Department of Physics, Rutgers University, Newark 07102, New Jersey USA}
 \author{Saswata Roy}
 \affiliation{Department of Physics, Rutgers University, Newark 07102, New Jersey USA}
 \author{Neepa T. Maitra}
 \affiliation{Department of Physics, Rutgers University, Newark 07102, New Jersey USA}
 \email{neepa.maitra@rutgers.edu}
 \date{\today}

%non-adiabatic dynamical steps and peaks in TDDFT, we can put a more  discussion on the steps in general in TDDFT, distinction with stepscoming from fractional charge problems like in ionization which would be more an adiabatic feature, their distinction to ground-state steps, their appearance in non-interacting systems when the initial state isdifferent, etc etc. 

\begin{abstract}
The adiabatic approximation in time-dependent density functional theory (TDDFT) is known to give an incorrect pole structure in the quadratic response function, leading to unphysical divergences in excited state-to-state transition probabilities and hyperpolarizabilties.  We find the form of the exact quadratic response kernel and derive a practical and accurate approximation that cures the divergence. We demonstrate our results on excited state-to-state transition probabilities of a model system and of the LiH molecule. 
\end{abstract}
\maketitle

Until recently, quadratic response has received far less attention than linear response. Most response applications had involved properties related to the optical spectra of a molecule in equilibrium, while relatively few ventured into non-linear regime to gain access to properties such as two-photon absorption, sum-frequency generation, and hyperpolarizabilities which can be obtained from the quadratic response of the ground-state system~\cite{PapadopoulosBook,MukamelBook}. However, in the past few decades, non-linear optical processes have emerged as key in a number of applications, including optical data storage and switching, for examples.  Moreover, an increasingly relevant class of applications involve excited-state dynamics, where a molecule is initially photo-excited and coupled electron-ion motion ensues. Such applications inherently require the response of an excited state, appearing in the form of excited state-to-state transition amplitudes. These amplitudes also appear even without nuclear motion: when simulating the dynamics of a molecule in a non-perturbative laser field by expressing the wavefunction in a superposition of eigenstates, coupled by the laser field. 

Response theory offers a way to obtain these quantities by circumventing the expensive calculation of the excited-state wavefunctions, and may yield more accurate properties when, inevitably, approximations are used. However, response theories of approximate electronic structure theories suffer from an unphysical divergence problem when the difference between two excitation frequencies is equal to another excitation frequency~\cite{PRF16}. This had been first discovered in time-dependent Hartree-Fock (TDHF) forty years ago~\cite{D82} but lay relatively dormant until the work of Ref.~\cite{PRF16} which showed the divergence also appears in  response theories based on coupled-cluster, multi-configuration self-consistent field, and in adiabatic time-dependent density functional theory (TDDFT)~\cite{LL14,OBFS15,ZH15, PRF16}. 

Addressing this issue for TDDFT~\cite{RG84,Carstenbook,TDDFTbook2012,M16} is of great interest: not only does TDDFT have a favorable system-size scaling enabling the calculation of photo-induced dynamics in complex molecules,
% yielding an unprecedented balance between accuracy and efficiency in the linear response regime, 
it is in principle an exact theory and so offers the possibility of finding more accurate functional approximations that cure the unphysical divergence, which is what we aim to achieve here.

We find the form of the exact quadratic response kernel of TDDFT and show explicitly why the adiabatic approximations used thus far are responsible for the incorrect pole structure of the second-order response function that creates the divergence, and that a relatively gentle linear frequency-dependence in the quadratic response kernel corrects the pole structure and tames the divergence. Inspired by this, we derive a frequency-dependent approximation for the quadratic response kernel. Results on a two-electron model system and on the LiH molecule show that our approximation provides a practical and accurate fix to the problem of divergences in TDDFT quadratic response.

In TDDFT response theory, the central object at each order of response is a density-response function expressed in terms of response functions of the Kohn-Sham (KS) system, and exchange-correlation kernels~\cite{GDP96,TDDFTbook2012}. 
The linear density response function of the interacting system to an external perturbation $\delta v\ext(\br,t)$, $\chi(\br,\br',t - t') = \frac{\delta n(\br,t)}{\delta v\ext(\br't')} = -i \theta(t-t')\langle \Psi_0 \vert [\hat{n}(\br t),\hat{n}(\br' t')]\vert\Psi_0\rangle$, where $\theta(t-t')$ is the step function, has the spectral representation 
\ben
\chi(\br,\br',\omega) = \sum_a \left(\frac{n_{0a}(\br)n_{a0}(\br')}{\omega - \Omega_a + i0^+} \, - \, \frac{n_{0a}(\br')n_{a0}(\br)}{\omega + \Omega_a + i0^+} \right)
\een
where
$n_{0a}(\br) = \langle\Psi_0\vert \hat{n}(\br) \vert \Psi_a\rangle$ is the transition density between the ground state, $\Psi_0$ and the excited state, $\Psi_a$ which has excitation frequency $\Omega_a = E_a - E_0$ and $\hat{n}(\br)$ is the one-body density-operator; the $0^+$ indicates the shift of the pole slightly below the real-axis to ensure causality and will be omitted hereon.
In TDDFT, $\chi$ is instead obtained from the non-interacting KS system, through the Dyson-like equation~\cite{PGG96,C95}
\ben
\chi^{\rm tddft}_{ij}(\omega) = \chi_{{\sss S},ij}(\omega) + \chi_{{\sss S},ik}(\omega)f_{\Hxc,kl}(\omega)\chi_{lj}(\omega)
\een
where  $\chi_{{\sss S}}$ is the density response function of the KS system and $f\Hxc(\omega)$ is the Hartree-exchange correlation kernel. The indices $i, j$ represent the spatial variables $\br_i$ and $\br_j$ and repeated indices imply integration. 
 While $\chi\s(\omega)$ displays residues given by transition-densities between ground and excited states of the KS system, and poles given by KS excitation frequencies, the  linear response (LR) kernel,  $f_{\Hxc,kl}[n_0](t - t') =\frac{\delta(t - t')}{\vert \br_k - \br_l\vert} + \left.\frac{\delta v\xc[n](\br_k,t)}{\delta n(\br_l,t')}\right\vert_{n = n_0}$, corrects these  to those of the true response function. 
Almost always, an adiabatic approximation is used, where the exchange-correlation potential $v\xc[n](\br,t)$ depends only on the instantaneous density and is approximated by the functional derivative of a ground-state energy functional, $E\xc[n]$. This results in a frequency-independent kernel, 
 %Adiabatic approximation, which is derived by taking the ground-state functionals and evaluating them on instantaneous density is the most popular way of constructing a frequency-independent approximation for the exchange-correlation functional, 
$f_{{\sss XC},ij}^{\rm adia}[n](\omega) = \frac{\delta^2E\xc[n]}{\delta n(\br_i) \delta n(\br_j)}
$. With an adiabatic approximation,  LR TDDFT has become a workhorse of electronic structure calculations, yielding excitation spectra with an unprecedented balance between accuracy and efficiency. The adiabatic approximation is known to fail for certain classes of excitations, and improved, frequency-dependent, approximations have been derived for some cases, e.g. double-excitations~\cite{MZCB04,M22}. 
 
 Going to second-order in the perturbation, defines the quadratic response (QR) function~\cite{WH74,GDP96,SS87}, $\chi^{(2)}(\br,\br_1,\br_2,t-t_1, t-t_2) =\frac{\delta^2 n(\br,t)}{\delta v\ext(\br_1,t_1)\delta v\ext(\br_2,t_2)} $: 
 \begin{multline}
 %\begin{equation}
\chi^{(2)}(\br,\br_{1},\br_{2};t-t_{1},t-t_{2})=\frac{(-i)^2}{2}\theta(t-t_{1})\theta(t_{1}-t_{2})\\
\times\langle\Psi_{0}\vert\left[\left[\hat{n}(\br,t),\hat{n}(\br_{1},t_{1})\right],\hat{n}(\br_{2},t_{2})\right]\vert\Psi_{0}\rangle \, +\,\,(1\leftrightarrow 2)
 %\end{equation}
 \end{multline}
which has the spectral representation~\cite{SS87}
\begin{widetext}
\bea
%\chi^{(2)}(\br,\br_i,\br_j,\omega_1,\omega_2) = \frac{1}{2}\sum_{a,b}\left( \frac{n_{0a}(\br)n_{ab}(\br_i)n_{b0}(\br_j)}{(\omega_1 + \omega_2 - \Omega_b)(\omega_2 - \Omega_a)} - \frac{n_{0a}(\br_j)n_{ab}(\br)n_{b0}(\br_i)}{(\omega_1 - \Omega_b)(\omega_2 + \Omega_a)} + \frac{n_{0a}(\br_j)n_{ab}(\br_i)n_{b0}(\br)}{(\omega_1 + \omega_2 + \Omega_b)(\omega_1 + \Omega_a)} + \, (i \leftrightarrow j)  \right)  
\chi^{(2)}(\br,\br_i,\br_j,\omega_i,\omega_j) = \frac{1}{2}\sum_{a,b}\left( \frac{n_{0a}(\br)n_{ab}(\br_i)n_{b0}(\br_j)}{(\omega_i + \omega_j - \Omega_b)(\omega_j - \Omega_a)} - \frac{n_{0a}(\br_j)n_{ab}(\br)n_{b0}(\br_i)}{(\omega_i - \Omega_b)(\omega_j + \Omega_a)} + \frac{n_{0a}(\br_j)n_{ab}(\br_i)n_{b0}(\br)}{(\omega_i + \omega_j + \Omega_b)(\omega_i + \Omega_a)} + \, (i \leftrightarrow j)  
\right)  
\label{eq:chi2}
\eea
where the state-$a$ to state-$b$ transition density is
$n_{ab}(\br) = \langle\Psi_a\vert \hat{n}(\br) \vert \Psi_b\rangle$, and can be extracted from double residues of $\chi^{(2)}$.  
The second-order response may be extracted from TDDFT linear response quantities together with a QR kernel $g\xc(\br,\br_{1},\br_{2},t-t_{1},t-t_{2})=\left.\frac{\delta^{2}v\xc(\br,t)}{\delta n(\br_1,t_1)\delta n(\br_2,t_2)}\right\vert_{n_{0}}$ through~\cite{GDP96,PF17,SVHA02}:
\begin{multline}
\chi_{mnp}^{(2),{\rm tddft}}(\omega_{1},\omega_{2})  =\chi_{mi}(\omega_1+\omega_2)\chi_{{\sss S},ij}^{-1}(\omega_{1}+\omega_{2})\chi_{{\sss S},jkl}^{(2)}(\omega_{1},\omega_{2})\chi_{{\sss S},lq}^{-1}(\omega_1)\chi_{qn}(\omega_1)\chi_{{\sss S},kr}^{-1}(\omega_2)\chi_{rp}(\omega_2)\\
  +\chi_{mi}(\omega_1+\omega_2)g_{{\sss XC},ijk}(\omega_1,\omega_2)\chi_{jn}(\omega_1)\chi_{kp}(\omega_2)
\label{eq:chi2-tddft}
\end{multline}
\end{widetext}
(again using the index notation for spatial dependences). In the adiabatic approximation, $g^{\rm adia}_{{\sss XC},ijk}(\omega_1,\omega_2) = \left.\frac{\delta^3 E\xc[n]}{\delta n_i \delta n_j \delta n_k}\right\vert_{n = n_0}$ is frequency-independent. 

Eq.~(\ref{eq:chi2-tddft}) is usually recast in terms of a matrix in the space of KS single-excitations in molecular codes, e.g.~\cite{dalton,turbomole2}, or written in a Sternheimer formulation~\cite{GV89,TDDFTbook2012}, 
which has enabled calculations of a wide range of non-linear optical properties of complex systems, e.g.~\cite{GSB97,ZWWFS21,NBJO05,KJOCH08,ZG14}
%, including second-harmonic generation, and excited-state transition matrix elements. 
However, several works encountered greatly exaggerated responses in domains where the difference between two excitation frequencies $\Omega_{b}$ and $\Omega_{c}$ is equal to another excitation frequency, $\Omega_{a}$, i.e. $\Omega_{c} - \Omega_{b} = \Omega_{a}$~\cite{HAJ16,PRF16,LSL14,OBFS15,ZH15}, which,  in this work, we call the ``resonance condition". 
Ref.~\cite{PRF16} tracked this unphysical divergence 
%adiabatic TDDFT shows an unphysical divergence when the difference between two excitation frequencies is equal to another excitation frequency, $\Omega_{c} - \Omega_{b} = \Omega_{a}$, which, in this work, we call the ``resonance condition". This can be tracked back 
to an incorrect pole structure in $\chi^{(2),{\rm tddft}}$ when an adiabatic approximation is made, pointing out the similarity to the divergence observed in Ref.~\cite{D82} for TDHF, as well as  in other response theories. The question arises: Since TDDFT is in principle an exact theory, what is the structure of the exact QR kernel that cures this divergence? And can we build a practical approximation that inherits this behavior? 

\begin{figure}[h]
\includegraphics[width=0.5\textwidth]{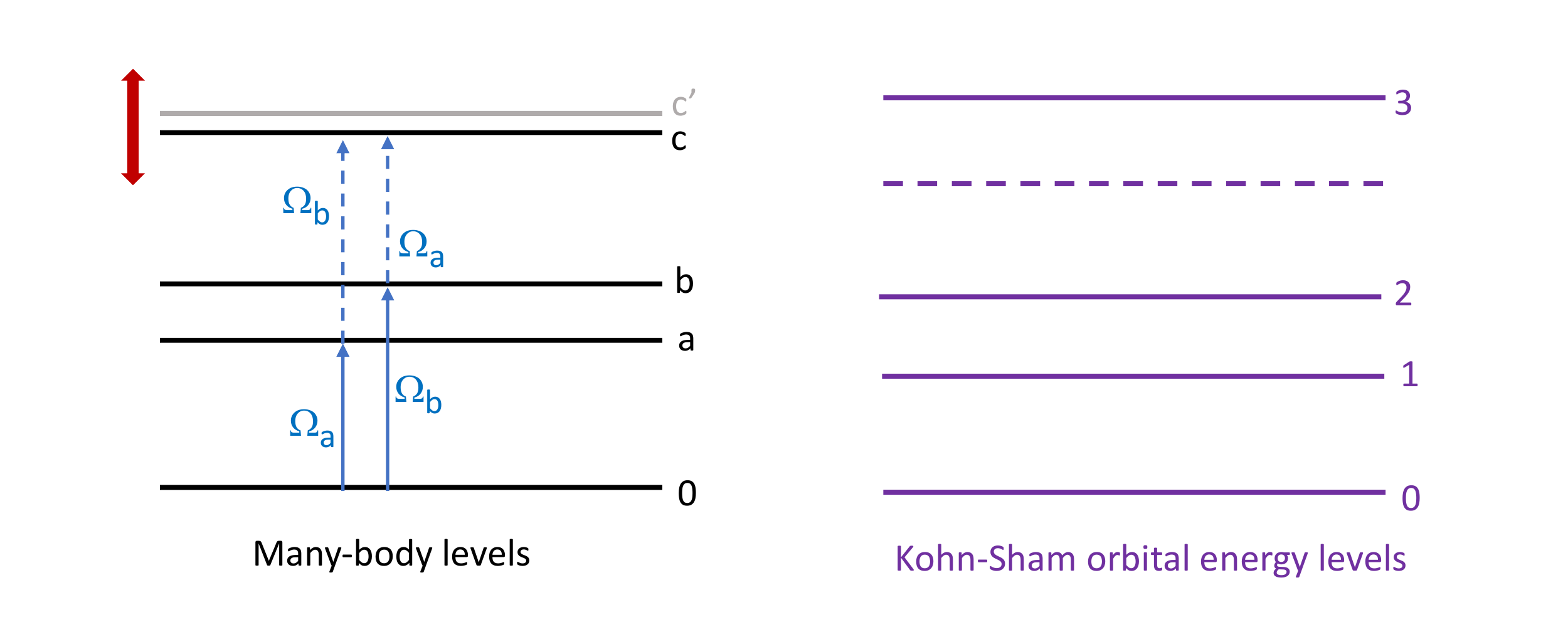} 
\caption{Depiction of a truncated Hilbert space. The true interacting system (left) with three excited states of excitation frequencies $\Omega_a,\Omega_b$ and $\Omega_{c}$, with  corresponding KS system (right) with excitation frequencies $\nu_1, \nu_2, \nu_3$ with 0 being the HOMO level; $\nu_i = \epsilon_i - \epsilon_0$ where $\epsilon_i$ is the orbital energy. The possibility of a KS double-excitation of frequency $\nu_1 + \nu_2$ (dashed line) mixing with the single excitation at $\nu_3$ yields an additional excitation $\Omega_c'$ (grey on the left); in this case $c$ and $c'$ have both single and double character. Displacement along the red vertical arrow tunes the system out of the resonance condition. 
In our model, we take the case when $\Omega_b = \Omega_a$ and $\nu_2 =\nu_1$. 
}
\label{fig_trunc}
\end{figure}

%Analysis of poles of density response with adiabatic QR has been carried out before \cite{EGCM11} and led to many insights. Inspired by this, 
To answer these questions, we  construct the exact $\chi^{(2)}$ in a Hilbert space truncated to contain four many-body states, denoted $0, a, c, c'$, and solve for the exact form of the QR kernel in this truncated space by inversion of Eq.~(\ref{eq:chi2-tddft}). The resonant case is met when $\Omega_c = 2\Omega_a$.  We include the possibility of a double-excitation contribution to the many-body states, where the states $c$ and $c'$ are approximately linear combinations of a single KS excitation $\nu_3$ and a double KS excitation $2\nu_1$ (see Fig.~\ref{fig_trunc} for a slightly more general truncation, and note that in this paper we will refer to KS excitations via the symbol $\nu$ and true excitations via the symbol $\Omega$).
In fact the resonance condition is suggestive of a state of double-excitation character: $\Omega_c = \Omega_b + \Omega_a$ would have double-excitation character if $\Omega_{a}$ and $\Omega_{b} $ are predominantly single excitations out of a Slater determinant reference and if the TDDFT corrections to the excited state energies are small.
%both the diagonal correction as well mixing with other KS excitations through the TD linear response function can shift the true excitations such that the resonance condition is far from the position of a double excitation. A double-excitation is involved in the first example below while single excitations dominate in the second example. 
 We will consider the second-order response at frequencies, $\omega_1, \omega_2$ that are much closer to $\Omega_a$ than to $\Omega_c$.

To simplify the inversion, we assume that the KS states have frequencies $\nu_1, \nu_3$ well-separated from each other, and far enough from the ground-state, such that the single-pole approximation may be applied~\cite{PGG96,GPG00,AGB03,TDDFTbook2012}. Then, constructing the linear response functions $\chi$ and $\chi\s$, and the KS quadratic response function $\chi\s^{(2)}$ and using them in Eq.~(\ref{eq:chi2-tddft})(see Supplemental Material for detail) gives
\begin{widetext}
\bea
\nonumber
\chi^{(2),{\rm tddft}}_{mnp}(\omega_1, \omega_2) &=& \left(\frac{a_{c,mi}}{\omega_1 +\omega_2 - \Omega_c} +  \frac{a_{c',mi}}{\omega_1 +\omega_2 - \Omega_{c'}} \right)\left[a_{{\sss S}3,ij}^{-1} \,\frac{\nu_1^2}{2\Omega_a^2} \left(\frac{A_{{\sss S}13, jnp}}{\omega_2 - \Omega_a} + \frac{A_{{\sss S}13, jpn}}{\omega_1 - \Omega_a} \right) \right. \\
&+& \left. \frac{a_{{\sss S}3,ij}^{-1}\,\frac{\nu_1^2}{\Omega_a^2} \langle f\Hxc\rangle_1 (A_{{\sss S}13, jnp} + A_{{\sss S}13, jpn}) + g_{{\sss XC},ijk}(\omega_1,\omega_2)a_{a,jn} a_{a,kp} }{(\omega_2 - \Omega_a)(\omega_1 - \Omega_a)} \right]
\label{eq:tddftchi2_1}
\eea
\end{widetext}
where $A_{\sss S13,jnp}=n_{\sss S01}(\br_j)n_{\sss S13}(\br_n)n_{\sss S30}(\br_p)$, $a_{\sss S3,ij}=n_{\sss S03} (\br_i)n_{\sss S30}(\br_j)$, defined in terms of the KS transition densities. The residue $a_{c,mi} = n_{0c}(\br_m)n_{c0}(\br_i)$ and $\langle f\Hxc(\omega_2)\rangle_1 = \int \phi_0(\br)\phi_1(\br) f\Hxc(\br,\br',\omega)\phi_0(\br')\phi_1(\br') d^3\br d^3\br'$. 
All quantities on the right of Eq.~(\ref{eq:tddftchi2_1}) can be obtained from LR TDDFT, the QR kernel, or from the KS system directly.

When $g\xc$ is independent of frequency, the incorrect pole structure is salient, with the last line of Eq.~(\ref{eq:tddftchi2_1}) having a three-pole structure instead of the two appearing in the exact $\chi^{(2)}$ of Eq.~(\ref{eq:chi2})~\cite{PRF16,LSL14,OBFS15}.
We note that in most cases the frequencies $\omega_{1,2}$ are in a region dominated by single excitations, where the adiabatic approximation for the linear response xc kernel $f\xc$ does a reasonable job, i.e. the exact $f\xc$ does not have important frequency-dependence in the region it is probed in Eq.~(\ref{eq:tddftchi2_1}). 
%To see this, we consider a system for which the resonance condition is satisfied at some configuration. We illustrate this with the picture in Fig.~\ref{hilbert_trun}. Without loss of generality, the external frequencies, $\omega_1$ and $\omega_2$ that probe the system are both taken to be roughly in the region of the first excited state $\Omega_a$. With these specifications, the double residue of the QR response function takes the form,
%\begin{widetext}
%\begin{multline}
% \begin{flalign}
% \ben
%\lim_{(\omega_2,\omega_1) \to (\Omega_a,\Omega_b-\Omega_a)}(\omega_2-\Omega_a)(\omega_1+\omega_2-\Omega_b)\chi_{mnp}^{(2),{\rm tddft}}(\omega_1,\omega_2)
%=a_{b,mi}\left[a^{-1}_{{\sss S} 2,ij}\frac{\nu^2_1}{2\Omega^2_a}A_{{\sss S}12,jnp} +a^{-1}_{{\sss S},ij}\frac{\nu^2_1}{2\Omega^2_a} \lim_{\omega_2 \to \Omega_a}\left(\frac{\omega_2-\Omega_a}{\omega_1-\Omega_a}\right)\right. &\\
%&+\left.\frac{a^{-1}_{{\sss S},ij}\frac{\nu^2_1}{\Omega^2_a}\langle f_{\Hxc}(\omega_2)\rangle_a (A_{{\sss S},jnp}+A_{{\sss S},jpn})+g_{\xc,ijk}(\omega_1,\omega_2)a_{a,jn}a_{a,kp}}{(\omega_1-\Omega_a)}\right] 
% \een
% \end{flalign}
%\end{multline}
%\end{widetext}
Instead, it follows that $g\xc$ must carry a frequency-dependence that removes the extra pole, which means the numerator of the last term in  Eq.~\ref{eq:tddftchi2_1} must be of the form:
$X_{inp} (\omega_1 - \Omega_a) + Y_{inp} (\omega_2 - \Omega_a)$ where $X_{inp}$ and $Y_{inp}$ are functions of $\{\br_i,\br_n,\br_p\}$.
The permutation-symmetry of $\chi^{(2)}$ under $(\br_1,\omega_1) \leftrightarrow (\br_2, \omega_2)$ implies $Y_{inp} = X_{ipn}$, leading to:
\begin{widetext}
\ben
X_{inp}(\omega_1 - \Omega_a)+ X_{ipn}(\omega_2 -\Omega_a) =g_{{\sss XC},ijk} (\omega_1, \omega_2) a_{a,jn} a_{a,kp}
+a_{{\sss S}3,ij}^{-1} \, \frac{\nu_1^2}{\Omega_a^2}  \langle f\Hxc(\omega_2) \rangle_1 (A_{{\sss S}13, jnp} + A_{{\sss S}13, jpn})
\label{eq:XY_eq}
\een
\end{widetext}
Eq.~(\ref{eq:XY_eq}) shows that the exact QR kernel $g\xc$ in the vicinity of $\omega_{1,2}$ close to $\Omega_a$ has a linear frequency-dependence. 
For the general case where $\Omega_a \neq \Omega_b$, a similar analysis leads to $g\xc(\omega_1 \approx \Omega_a, \omega_2\approx \Omega_b)$ having a linear behavior as $X(\omega_1 - \Omega_a) + Y (\omega_2 - \Omega_b)$.
It remains now to derive an approximation for $X_{inp}$ which yields a practical approximation for $g_{{\sss XC},ijk}(\omega_1,\omega_2)$.

In order to determine $X_{inp}$, we interpolate between two limiting cases. The first is to set $g\xc \to g\xc^{\rm adia}$ when $\omega_1 = \omega_2 = 0$ in Eq.~(\ref{eq:XY_eq}), which gives an equation for $X_{inp} + X_{ipn}$. A possible solution is
{\small
\ben
%\nonumber
X_{inp} = \frac{-1}{2\Omega_a}\left(g_{{\sss XC},ijk}^{\rm adia} a_{a,jn} a_{a,kp}
%\right. \\\left.
+ a_{{\sss S}3,ij}^{-1}\, \frac{\nu_1^2}{\Omega_a^2}  \langle f\Hxc(\omega_2) \rangle_1  A_{{\sss S}13, jnp}\right)
\label{X_inp}
\een}
Using Eq.~(\ref{X_inp}) in Eq.~(\ref{eq:XY_eq}) gives $g\xc^{\rm App,1}(\omega_1,\omega_2)$ that corrects the single-excitation contribution ($A_{{\sss S}13, jnp}$ ) to the quadratic response but appears not to include double-excitation contributions to the transition density. It is unclear whether the first term captures true double-excitation character because an adiabatic QR kernel yields a response that has poles at sums of LR-corrected single excitations without any mixing with double-excitations and even these poles are missing when only forward transitions are kept~\cite{EGCM11,TC03}. Our second limiting case therefore focusses on the double-excitation contribution.

Thus the second limit is the opposite case when state $c$ is a close to a pure double excitation. Considering Fig.~\ref{fig_trunc} the KS state 3 is absent and we denote the  KS state with two electrons excited to orbital 1 at the dashed line, as $d$. The KS residue appearing in Eq.~(\ref{eq:tddftchi2_1}),  $A_{{\sss S}1d,ijk} = n_{{\sss S}01}(\br_i)n_{{\sss S}1d}(\br_j)n_{{\sss S}d0}(\br_k) = 0$ due to the last factor, and equating Eq.~(\ref{eq:tddftchi2_1}) to the true $\chi^{(2)}$ in this limit yields
%\begin{widetext}
\bea
%(a_bg^{\rm App-2}\xc a a)_{mnp} = \frac{1}{2}\left[\left(\omega_1 - \Omega_a\right)A_{ba,mnp} + \left(\omega_2 - \Omega_a\right)A_{ba,mpn}\right]
\nonumber
a_{c,mi}g^{\rm App,2}_{{\sss XC},ijk} a_{a,jn} a_{a,kp} &=&\frac{1}{2}\left[\left(\omega_1 - \Omega_a\right)A_{ca,mnp}\right. \\ &+& \left.\left(\omega_2 - \Omega_a\right)A_{ca,mpn}\right]
\label{near_d}
\eea
%\end{widetext}
The residue $A_{ca} = n_{0c}n_{ac}n_{a0}$, contains the ground-to-excited transition densities of the true system $n_{0a}$ and $n_{0c}$ which are accessible from LR, and
substituting the KS excited-to-excited transition density $n_{1d}$
%$ =\langle\Phi_1\vert\hat{n}(\br)\vert\Phi_d\rangle =\sqrt{2}\phi_0\phi_1$ 
for $n_{ac}$ in Eq.~(\ref{near_d}) gives the second limit in our approximation for $g\xc$. Our final approximation interpolates between the two limits through the weighting of the double-excitation component to the true state 
 (details in the Supplemental Material), 
\begin{widetext}
\bea
\label{eq:approx}
%\nonumber
%g_{{\sss XC},iqr}^{\rm App}(\omega_1,\omega_2) &=& -\left(\frac{\omega_1 + \omega_2 - 2 \Omega_a}{2\Omega_a}\right)g_{{\sss XC},iqr}^{\rm adia} - a_{{\sss S}2,ij}^{-1} \frac{\nu_1^2}{2\Omega_a^3} (\Omega_a - \nu_1)(\omega_1 A_{{\sss S}12, jnp} + \omega_2 A_{{\sss S}12, jpn})a_{a,nq}^{-1} a_{a,pr}^{-1} \\
%&+& 
%\frac{a^{-1}_{b,im}}{2}\sqrt{1 - a_b a^{-1}_{{\sss,S}3}}\frac{n_{0b}(r_m)}{2}\left[\left(\omega_1 - \Omega_a\right)n_{d1}(r_n)n_{0a}(r_p) + \left(\omega_2 - \Omega_a\right)n_{d1}(r_p)n_{0a}(r_n) \right]a_{a,nq}^{-1} a_{a,pr}^{-1}
g_{{\sss XC},iqr}^{\rm App}(\omega_1,\omega_2) &=& -\left(\frac{\omega_1 + \omega_2 - 2 \Omega_a}{2\Omega_a}\right)g_{{\sss XC},iqr}^{\rm adia} - \left(a_{{\sss S}3,ij}^{-1} \frac{\nu_1^2}{\Omega_a^3} \langle f\Hxc(\omega_2) \rangle_1 (\omega_1 A_{{\sss S}13, jnp} + \omega_2 A_{{\sss S}13, jpn})\right. \\
%\label{eq:approx}
\nonumber
&+& 
\left.\frac{a^{-1}_{c,im}}{2}(\sqrt{1 - a_c a^{-1}_{{\sss S}3}})_{mo}\frac{n_{0c}(r_o)}{2}\left[\left(\omega_1 - \Omega_a\right)n_{d1}(r_n)n_{0a}(r_p) + \left(\omega_2 - \Omega_a\right)n_{d1}(r_p)n_{0a}(r_n) \right]\right)a_{a,nq}^{-1} a_{a,pr}^{-1}
%\label{eq:approx}
\eea
\end{widetext}
For an excited state $c$ that has  predominantly single-excitation character, the first two terms dominate, while the third term incorporates the effect of its doubly-excited character. As evident from Eq.~(\ref{eq:approx}), all ingredients for our approximation can be obtained from linear response TDDFT, or adiabatic QR TDDFT.
Turning to the transition density obtained from the double-residue
\ben
\xi n_{ca}(\br_m) = \lim_{\substack{\omega_2 \to \Omega_a\\ \omega_1 +\omega_2 \to \Omega_c}} (\omega_2 - \Omega_a)(\omega_1+ \omega_2 - \Omega_c)\frac{\chi^{(2)}_{mmm}(\omega_1,\omega_2)}{n_{0c}(\br_m) n_{a0}(\br_m)}
\label{eq:nab}
\een
where $\xi = 1$ for the resonant case, and $\xi = \frac{1}{2}$ otherwise, we find 
\begin{widetext}
\ben
n_{ca}^{\rm App}(\br) = \sqrt{1 - \alpha_c^2}n_{{\sss S}1d}(\br) + \left(\frac{\nu_1}{\Omega_a}\right)^{3/2}\alpha_c \left(1 - 2\frac{\langle f\Hxc(\omega_2) \rangle_1 }{\Omega_a}\right)n_{{\sss S},13}(\br) - \frac{n_{0a}(\br)}{\Omega_a}\int n_{0c}(r_1)g\xc^{\rm adia}(r_1,r_2,r_3) n_{0a}(\br_2)n_{0a}(\br_3)dr_1 dr_2 dr_3
\label{eq:approx-nac}
\een
\end{widetext}
Here $\alpha_c^2$ is an $\br$-independent approximation to $a_c a^{-1}_{{\sss S},3}$: $\alpha_c$ ranges from $\sqrt{\nu_3/\Omega_c}$ in the case where the state $c$ is predominantly a single excitation, to $0$ when it is predominantly a double-excitation.
Eq.~(\ref{eq:approx-nac})  can be compared with the adiabatic approximation, for which
\begin{widetext}
\ben
n_{ca}^{\rm adia}(\br) = \left(\frac{\nu_1}{\Omega_a}\right)^{3/2}\alpha_c \left(1 + \frac{\Omega_a - \nu_1}{\Omega_c - 2\Omega_a }\right) n_{\sss S13}(\br) + \frac{n_{0a}(\br)}{\Omega_c - 2\Omega_a}\int n_{0c}(r_1)g\xc^{\rm adia}(\br_1,\br_2,\br_3) n_{0a}(\br_2)n_{0a}(\br_3)dr_1 dr_2 dr_3\,.
\label{eq:adia-nac}
\een
\end{widetext}
The divergence is evident in the last two terms when the resonant condition, $\Omega_c - \Omega_a = \Omega_a$ is satisfied; further, there is no contribution from any double-excitation.

In practice, there are several {\it ad hoc} workarounds to the unphysical divergence, including applying damping factors~\cite{dalton}, neglecting the kernels in ``simplified-TDDFT"~\cite{BG14}, setting the term to zero, and the pseudo-wavefunction approximation~\cite{LL14,OAS15,AOS15,OBFS15,PRF16} where orbital relaxation terms in the second-order response are neglected, which is equivalent to solving the second-order response equation at zero frequency. 
The second term in Eq.~(\ref{eq:approx-nac}) could be viewed as in the pseudo-wavefunction spirit in the sense that it can be obtained by setting $\omega_1$ to zero in the divergent term of Eq.~(\ref{eq:adia-nac}). The connections with the standard pseudo-wavefunction approximation are left for future work, including what the implied underlying $g\xc$ kernel is; our work suggests it also has a linear frequency-dependence. In any case, all the standard workarounds miss any double-excitation contribution to the transition density (the first term in Eq.~(\ref{eq:approx-nac})), which can be significant as our first example below demonstrates.

%{\it i may just end up deleting this para -- depends on earlier part with description of the $\alpha_b$} We note that the resonance condition may be suggestive of a state of double-excitation character: $\Omega_c = \Omega_b + \Omega_a$ would have double-excitation character if $\Omega_{a,b}$ are predominantly single excitations out of a Slater determinant reference. However, this argument only holds in the limit that the TDDFT corrections to the excited state energies are small: both the diagonal correction as well mixing with other KS excitations through the TD linear response function can shift the true excitations such that the resonance condition is far from the position of a double excitation. A double-excitation is involved in the first example below while single excitations dominate in the second example. 

Our first example is a model system of two electrons in a one-dimensional harmonic plus linear potential, $v\ext(x) = \frac{1}{2}x^2 + \gamma \vert x\vert$ where $\gamma$ is a parameter in the range$[-1,1]$; varying $\gamma$ tunes the system in and out of the resonance condition. The electrons interact via a soft-Coulomb interaction: $\frac{\lambda}{\sqrt{(x_1-x_2)^2+1}}$, where we consider $\lambda = 0.2$a.u. as a weak interaction in which the assumptions made in Eqs.~(\ref{eq:tddftchi2_1})--(\ref{eq:approx-nac}) apply, but we also consider the results at the full coupling strength $\lambda = 1$a.u. In order to test our approximation for the QR kernel alone, without conflating errors from approximations made to the LR treatment,  we will use the {\it exact} KS and LR quantities in the equations. The LR thus includes double-excitation contributions, which would be missing in an adiabatic LR treatment. 

\begin{figure}[h]
\includegraphics[width=0.5\textwidth]{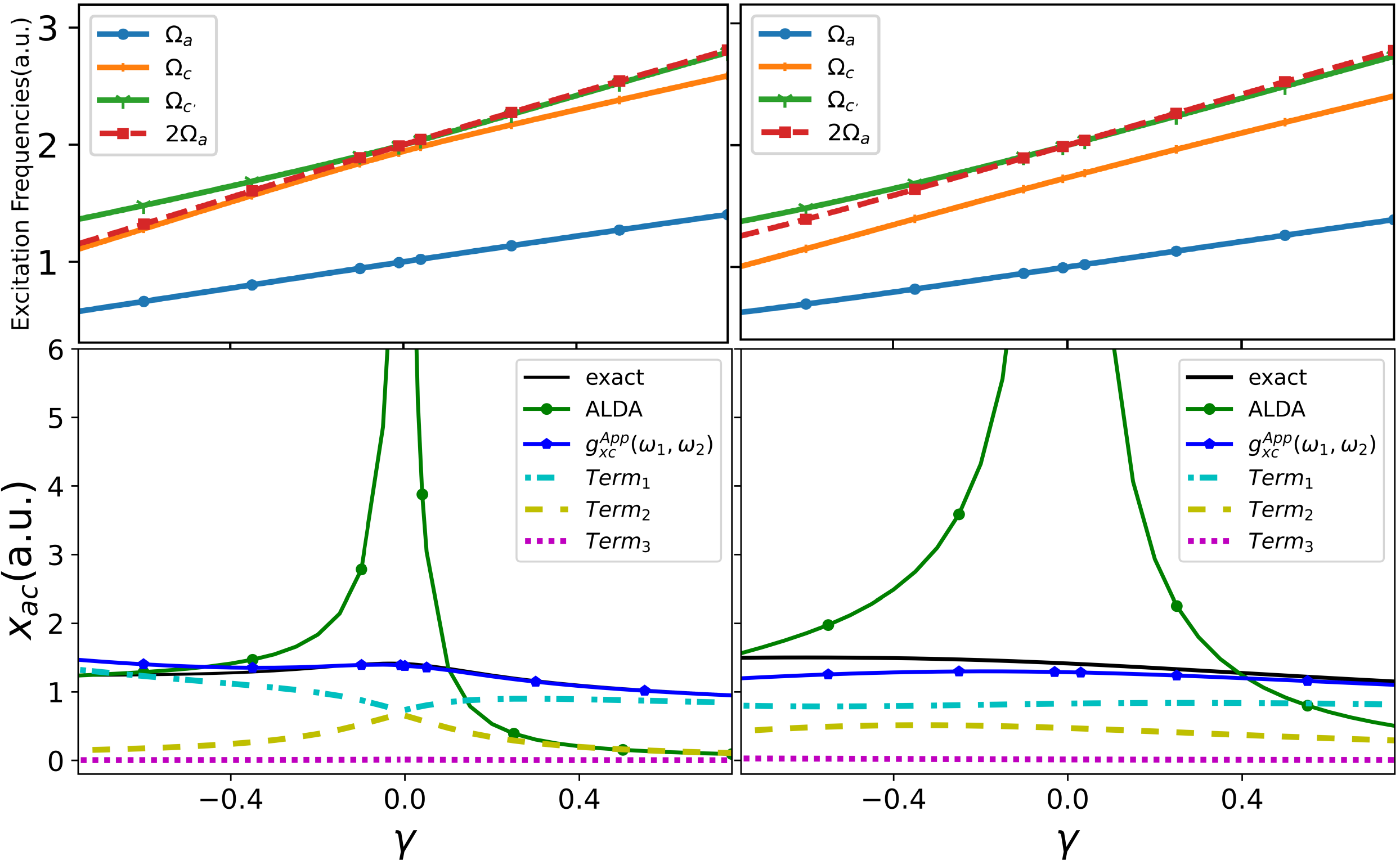} 
\caption{Excitation frequencies (upper) and transition dipole moments (lower panels), $x_{ac} = \langle\Psi_a\vert \hat{x} \vert \Psi_c\rangle$  between the first and 2nd excited states of the model system as a function of parameter, $\gamma$, for $\lambda=0.2$ (left) and $\lambda=1$ (right): exact (black), ALDA (green), our  approximation Eq.~(\ref{eq:approx-nac}) (blue). The dashed lines term 1, term 2, and term 3 are the respective terms of Eq.~(\ref{eq:approx-nac})}
\label{modelfig}
\end{figure}

Fig.~\ref{modelfig} shows the transition dipole moment, $\mu_{ac} = \langle\Psi_a\vert \hat{x} \vert \Psi_c\rangle$ between the first two excited states, $\Psi_a$ and $\Psi_c$ for which the resonance condition, in this case $\Omega_c=2\Omega_a$,  holds as $\vert\gamma\vert \rightarrow 0$. We calculate $\alpha_c$ from the ratio of the matrix element $\langle \Psi_0 \vert \hat{x}^2 \vert \Psi_c\rangle$ to the corresponding matrix element of the KS system. At the weaker coupling strength $\lambda=0.2$, our approximation Eq.~(\ref{eq:approx-nac}) 
 clearly cures the divergence of ALDA shown, and is barely distinguishable from the exact result in quite a wide region around the divergence.  Tuning  $\gamma$ away from $\gamma=0$, we move away from the resonance condition, and eventually we expect that our approximation may decrease in quality compared to ALDA: the error in our approximation from neglecting the mixing of other single excitations may no longer be negligible compared to the large error caused by the spurious pole in adiabatic approximations in the resonance region. For the particular case here, our approximation continues to do well for positive $\gamma$ where the system remains harmonic at large distances, while for negative values a double-well develops in $v\ext$ which brings the two lowest energy levels closer together as delocalized orbitals, deviating from the more clearly separated levels of a single well, and leading to a breakdown of the single-pole-like assumptions in the derivation of Eq.~(\ref{eq:approx-nac}).
Although our approximation was derived in the limit of well-separated excitations, we still observe a good performance at full coupling strength $\lambda=1.0$ (right panel in Fig.~\ref{modelfig}), not only curing the divergence seen in ALDA but also giving predictions close to the exact.

\begin{figure}[h]
\includegraphics[width=0.5\textwidth]{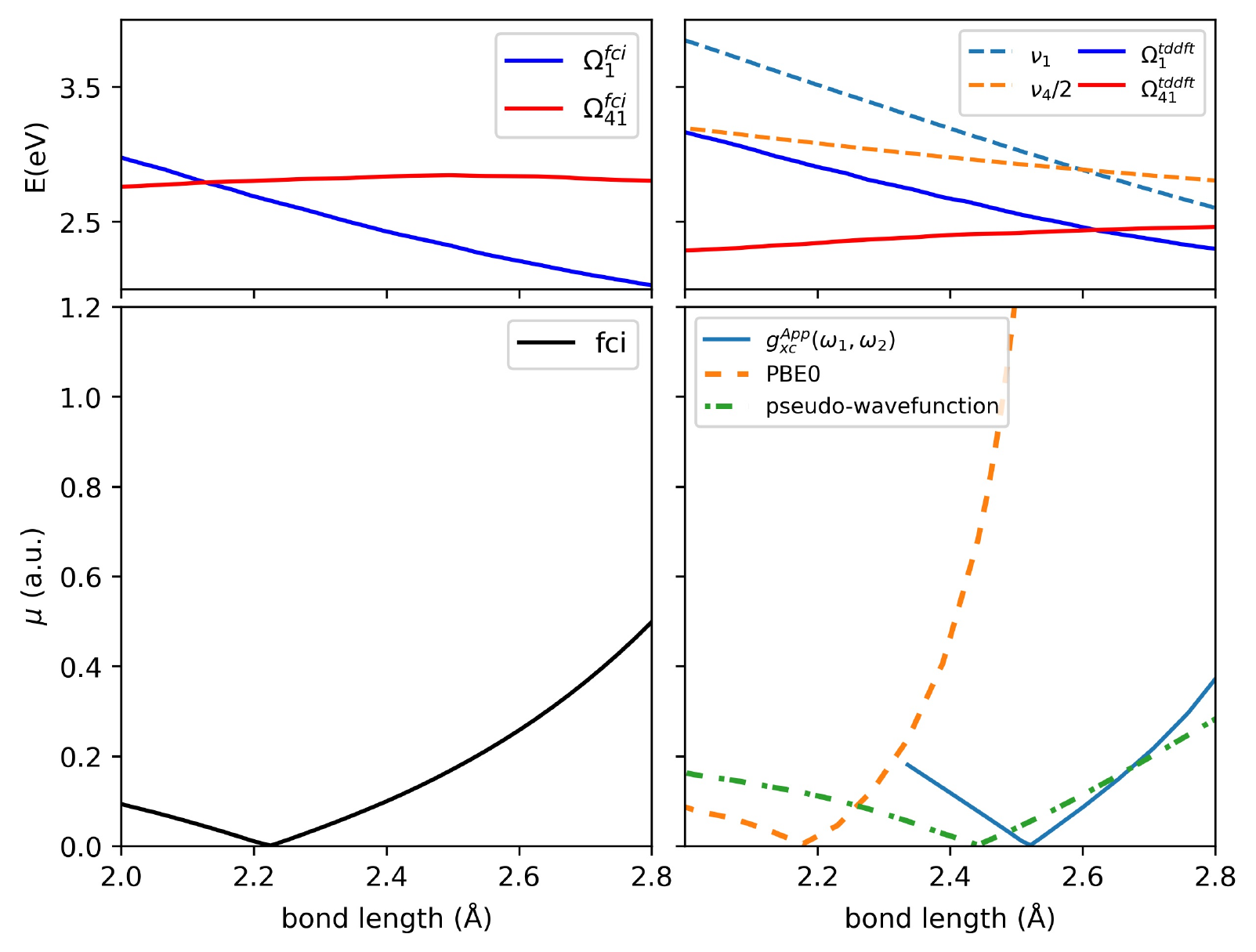} 
\caption{Upper panel: (left) Full configuration interaction (fci) and (right) TDDFT frequencies between the ground and the first excited state (solid blue), and the first and fourth excited state of LiH, $\Omega_{41} = \Omega_4 - \Omega_1$(solid red). The right panel also shows the KS lowest transition frequency, $\nu_1$ (light blue dotted ) and half of the fourth KS frequency, $\nu_4$ (orange dotted). Lower panel: Transition dipole moment, $\mu_{14}$  between the first and fourth excited states as a function of bond-length; exact (black), our approximation Eq.~(\ref{eq:approx-nac})(light blue), PBE0 (orange dashed), and pseudo-wavefunction approximation (green dashed). Note that for bond-lengths greater than $2.4 \AA$ the adiabatic PBE0 is out of the scale of the figure.} 
\label{LiHfig}
\end{figure}

We now turn to LiH, using PBE0~\cite{PBE0} with def2-SVP basis set~\cite{SVP}, within the Turbomole package~\cite{turbomole2}. The fourth excited state frequency is close to twice the first excited state frequency in the region around 2.6$\AA$ in PBE0; the first and fourth excited states correspond to the first  two excited states in the A$_1$ irreducible representation of $C_{2v}$ point group symmetry.  Also around this bond-length, the frequency of the lowest doubly-excited KS state matches that of the 4th single KS excitation (see top right panel Fig.~\ref{LiHfig}).  The PBE0 excitation energies are a little shifted from those of the reference full configuration interaction calculation taken from Ref.~\cite{PRF16}, while the transition dipole moment diverges in the resonance region $\Omega_4 = 2\Omega_1$~\cite{PRF16}. Our approximation, applied together with adiabatic PBE0 LR, shown in Fig.~\ref{LiHfig} tames this divergence, and follows the trend of the exact result, but with an overestimate; the adiabatic LR lacks the double-excitation contribution which, from the upper right panel could be expected to be significant, and so only the second term in Eq.~(\ref{eq:approx-nac}) contributes. Further, we set the third term to zero, because  we ran into some numerical problems in its extraction from Turbomole; we note that for the two-electron case of the harmonic oscillator where $g\x^{\rm adia}$ is strictly zero and $g\c^{adia}$ was small, it may be small in this case as well.
Likely, including this together with the double-excitation contribution with the LR kernel of Ref.~\cite{MZCB04} should improve the performance of our QR kernel; we leave for future work the investigation of oscillator strengths from a modified dressed LR TDDFT~\cite{MW09,C95,CFMB18} which would be used to determine $\alpha_c$. 
The figure shows also the result from the pseudo-wavefunction approximation that is often used~\cite{PRF16,AOS15,OBFS15}, which, despite being an ad hoc correction, appears to perform a little better than ours on  when compared with the same relative position in FCI.
 Again, as we move away from the resonance region, our approximation deviates as expected; the prescription would be that we revert to adiabatic PBE0 when the curves meet.

In summary, we found the form of the exact frequency-dependent kernel in QR TDDFT and  derived an approximate kernel based on this. Tests on the excited-state transition amplitudes of a model system and on the LiH molecule suggest it is a promising practical cure to the unphysical divergence problem in adiabatic QR TDDFT; Eq.~(\ref{eq:approx}) can be applied also to cure the divergences in other second-order response properties such as hyperpolarizabilities, two-photon absorption, etc. Our approach can be generalized to situations beyond the single-pole-type of analysis here, when more than one KS single-excitation contributes to a given state.  Alternatives to the two limiting approximations that we interpolated between here may lead to improved accuracy and will be explored in future work. This work also stresses the importance of including double-excitation contributions in LR; the kernel of Ref.~\cite{MZCB04} needs to be generalized to describe how the oscillator strength gets redistributed in this case~\cite{C95,MW09,CFMB18}. We note that TDDFT can also be applied in the real-time domain to obtain non-linear optical properties~\cite{CPRMR14} and the implications of the frequency-domain divergences for the time-domain have yet to be explored. Future work also includes determining whether the divergence is related to the spurious pole shift of generalized LR TDDFT~\cite{FLSM15,LFM16, M16} in the adiabatic approximation, as was previously surmised~\cite{PRF16}. The QR kernel is the functional derivative, or response, of the LR kernel evaluated at the ground-state,  and in this sense may be viewed as containing information about the linear response of an excited state. On the other hand, the pole shifting occurs quite generally, not just in situations where the resonance condition is satisfied. 
\section{Supporting Information } 
Detailed derivations and discussion of Eq (6) and the steps in the approximation leading to Eq. (10).

%{\it {\color{blue} wait, should the 1 be  $\nu_3/\Omega_c$ or somehow the factor involves the osc strength split instead???}}

\acknowledgments{
Financial support from the National Science Foundation Award CHE-1940333 (NTM), CHE-2154929 (DD),  and from the Department
of Energy, Office of Basic Energy Sciences, Division of Chemical
Sciences, Geosciences and Biosciences under Award No. DESC0020044 (SR) is gratefully acknowledged. Supplement funding for this project was provided by the Rutgers University at Newark Chancellor's Research Office. }
\onecolumngrid
\newpage

\renewcommand{\theequation}{S.\arabic{equation}}

\subsection{ (I)~Derivation of the TDDFT Second Order Response Equation in the Truncated Hilbert Space}
The Dyson equation for the second-order response function was derived in Ref.~\cite{GDP96} in terms of the linear response xc kernel, the KS response function, and the quadratic response xc kernel. (We note a typo in Eq. (173) of Ref.~\cite{GDP96} where the step functions on the right should read $\theta(t - \tau)\theta(\tau - \tau')$). To transform the time-domain expression into the frequency-domain, we Fourier transform using the factors $e^{i\omega_1(t-t_1)}e^{i\omega_1(t-t_2)}$~\cite{EGCM11}. Combining the resulting equation with the Dyson equation in linear response, we obtain 
{\small
\bea
\nonumber
\chi_{mnp}^{(2),{\rm tddft}}(\omega_{1},\omega_{2}) & =&\chi_{mi}(\omega_1+\omega_2)\chi_{{\sss S},ij}^{-1}(\omega_{1}+\omega_{2})\chi_{{\sss S},jkl}^{(2)}(\omega_{1},\omega_{2})\chi_{{\sss S},lq}^{-1}(\omega_1)\chi_{qn}(\omega_1)\chi_{{\sss S},kr}^{-1}(\omega_2)\chi_{rp}(\omega_2)\\
 & +&\chi_{mi}(\omega_1+\omega_2)g_{{\sss XC},ijk}(\omega_1,\omega_2)\chi_{jn}(\omega_1)\chi_{kp}(\omega_2)\,,
\label{eq:s_chi2-tddft}
\eea}
where, as in the main text, the subscripts represent the spatial variables.
 
We will evaluate the terms explicitly within the assumption of the truncated subspace discussed in the main text. The frequencies $\omega_1, \omega_2$ are taken to be far closer to $\Omega_a$ than to $\Omega_c$, while their sum $\omega_1+\omega_2$ is far closer to $\Omega_c$ than to $\Omega_a$. Further, we make a single-pole (Tamm-Dancoff)-like approximation in that we neglect the backward transitions, but we restore the oscillator strength sum-rule when relating the transition amplitudes of the KS and interacting systems.

With these considerations, we find the first-order response functions: 
\ben
\chi_{ij}(\omega_{1(2)}) \approx \frac{a_{a,ij}}{\omega_{1(2)} - \Omega_a} \;\;\; {\rm where} \;\;\;  a_{a,ij} = n_{0a}(\br_i)n_{a0}(\br_j), \;\;\;{\rm with} \;\;\; n_{0a}(\br) = \langle\Psi_0\vert \hat{n}(\br)\vert \Psi_a\rangle
%+ \frac{a_{c,ij}}{\omega_1 - \Omega_c} + \frac{a_{c',ij}}{\omega_1 - \Omega_{c'}} \approx \frac{a_{c,ij}}{\omega_1 - \Omega_c}
\een

\bea
\nonumber
\chi_{{\sss S},ij}(\omega_{1(2)}) \approx \frac{a_{{\sss S}1,ij}}{\omega_{1(2)} - \nu_1}, \;\;\; {\rm where} \;\;\; a_{{\sss S}1,ij} = n_{{\sss S}01}(\br_i)n_{{\sss S}10}(\br_j) \\
\eea
where the subscript ${\sss S}$ indicates KS quantities, and  similarly, 
\ben
\chi_{ij}(\omega_1 + \omega_2) \approx \frac{a_{c,ij}}{\omega_1 + \omega_2 - \Omega_c} + \frac{a_{c',ij}}{\omega_1 + \omega_2 - \Omega_{c'}}
\een
\ben
\chi_{{\sss S} ij}(\omega_1 + \omega_2) \approx \frac{a_{{\sss S3},ij}}{\omega_1 + \omega_2 - \nu_3} 
\een
For the second order KS response function we have
\ben
\chi^{(2)}_{{\sss S},jkl}(\omega_1,\omega_2) \approx \frac{1}{2}\left(\frac{A_{{\sss S}13,jkl}}{(\omega_1 + \omega_2 - \nu_3)(\omega_2 - \nu_1)} + \frac{A_{{\sss S}13,jlk}}{(\omega_1 + \omega_2 - \nu_3)(\omega_1 - \nu_1)} \right) 
\een
We make use of the single-pole approximation  \ben
\Omega_a = \nu_1 + 2 \langle f\Hxc \rangle_1 \;\; {\rm where  } \;\; \langle f\Hxc \rangle_1 \equiv \int d\br d\br' \phi_0(\br)\phi_1(\br)f\Hxc(\br,\br',\omega)\phi_0(\br')\phi_1(\br')
\een
and impose the oscillator strength sum rule within the single-excitation approximation
\ben
\Omega_a a_a \approx \nu_1 a_{{\sss S1}}
\een
to express 
\ben
\chi_{{\sss S},kr}^{-1}(\omega_2)\chi_{rp}(\omega_2) \approx \frac{\omega_2 - \nu_1}{\omega_2 - \Omega_a}\frac{\nu_1}{\Omega_a}\delta_{kp}
\een
 We note that making use of the single-pole approximation includes a diagonal correction to the KS excitation frequency and neglects coupling of the single-excitation to other single-excitations through off-diagonal elements of the kernel; this is justified  by the stated assumption that the KS orbital energies are well-separated. This assumption could be relaxed however, at the cost of a more complicated expression for the kernel; this would be in the same spirit as was done in a different context, Ref.~\cite{CZMB04}, when applying the dressed TDDFT kernel to double-excitations in linear polyenes, for example. Since the residues of the response functions are of central interest in this work, we wish to respect and restore the oscillator strength sum-rule as above~\cite{C95,GPG00,AGB03}.

Putting these together we finally obtain
\bea
\nonumber
\chi^{(2),{\rm tddft}}_{mnp}(\omega_1, \omega_2) &=& \left(\frac{a_{c,mi}}{\omega_1 +\omega_2 - \Omega_c} +  \frac{a_{c',mi}}{\omega_1 +\omega_2 - \Omega_{c'}} \right)\left[a_{{\sss S}3,ij}^{-1} \,\frac{\nu_1^2}{2\Omega_a^2} \left(\frac{A_{{\sss S}13, jnp}}{\omega_2 - \Omega_a} + \frac{A_{{\sss S}13, jpn}}{\omega_1 - \Omega_a} \right) \right. \\
&+& \left. \frac{a_{{\sss S}3,ij}^{-1}\,\frac{\nu_1^2}{\Omega_a^2} \langle f\Hxc \rangle_1 (A_{{\sss S}13, jnp} + A_{{\sss S}13, jpn}) + g_{{\sss XC},ijk}(\omega_1,\omega_2)a_{a,jn} a_{a,kp} }{(\omega_2 - \Omega_a)(\omega_1 - \Omega_a)} \right]
\label{eq:s_tddftchi2_1}
\eea
which is Eq.~(6) in the main paper. 

\subsection{(II) Approximation for the QR Kernel}
As observed in the paper, an adiabatic approximation for $g\xc(\omega_1,\omega_2)$ leaves the expression for $\chi^{(2),{\rm tddft}}$ with an excess pole compared to the exact $\chi^{(2)}$, which is ultimately responsible for the divergence in the residues. An approximation for $g\xc$ must have a frequency dependence that removes this pole which means the numerator of the last term of Eq.~(\ref{eq:s_tddftchi2_1}) above must be of the form:
\ben
X_{inp} (\omega_1 - \Omega_a) + Y_{ipn} (\omega_2 - \Omega_a) 
\label{eq:s_XY}
\een 
Considering the  symmetry of $\chi^{(2)}$ under $(\br_n,\br_p,\omega_1) \leftrightarrow (\br_p,\br_n,\omega_2)$, we deduce that $Y_{ipn} = X_{ipn}$, and so the approximation for $g\xc(\omega_1,\omega_2)$ reduces to finding an approximation for $X_{inp}$. Equating the numerator of the last term of Eq.~(\ref{eq:s_tddftchi2_1}) to Eq.~(\ref{eq:s_XY}) with $Y_{ipn} = X_{ipn}$, gives Eq. (7) in the main text. 

As discussed in the main paper, our approximation is derived from two limiting cases for which we provide more detail here. 

(i) {\underline{\it First limiting case:}} We set $g\xc \rightarrow g\xc^{\rm adia}$ when $\omega_1 = \omega_2 = 0$ in Eq (7) of the main text, which gives Eq. 8 of the main text for $X_{inp}$. Such an approximation leads to the QR kernel 
\ben
g_{{\sss XC},iqr}^{\rm App-1}(\omega_1,\omega_2) = -\left(\frac{\omega_1 + \omega_2 - 2 \Omega_a}{2\Omega_a}\right)g_{{\sss XC},iqr}^{\rm adia} - a_{{\sss S}3,ij}^{-1} \frac{\nu_1^2}{\Omega_a^3} \langle f\Hxc(\omega_2)\rangle_1 (\omega_1 A_{{\sss S}13, jnp} + \omega_2 A_{{\sss S}13, jpn})a_{a,nq}^{-1} a_{a,pr}^{-1} 
\een
which, when inserted into Eq.(6), gives
{\small{
\bea
\nonumber
\chi^{(2),{\rm App-1}}_{mnp}(\omega_1, \omega_2) &=& \left(\frac{a_{c,mi}}{\omega_1 +\omega_2 - \Omega_c} +  \frac{a_{c',mi}}{\omega_1 +\omega_2 - \Omega_{c'}} \right)\left[a_{{\sss S}3,ij}^{-1} \,\frac{\nu_1^2}{2\Omega_a^2} \left(\frac{A_{{\sss S}13, jnp}}{\omega_2 - \Omega_a} + \frac{A_{{\sss S}13, jpn}}{\omega_1 - \Omega_a} \right)\left( 1 - \frac{2\langle f\Hxc(\omega_2)\rangle_1}{\Omega_a}\right)  \right. \\
%&-&\left. \frac{1}{2\Omega_a}\left(\frac{1}{\omega_1 - \Omega_a} + \frac{1}{\omega_2 - \Omega_a} \right)\left(g_{{\sss XC},ijk}^{\rm adia}a_{a,jn} a_{a,kp} + a_{{\sss S}2,ij}^{-1} \frac{\nu_1^2}{\Omega_a^2} \langle f\Hxc(\omega_2)\rangle_a (A_{{\sss S}12, jnp} + A_{{\sss S}12, jpn})\right) \right]
&-&\left. \frac{1}{2\Omega_a}\left(\frac{1}{\omega_1 - \Omega_a} + \frac{1}{\omega_2 - \Omega_a} \right)g_{{\sss XC},ijk}^{\rm adia}a_{a,jn} a_{a,kp}\right]
%\left(\frac{1}{\omega_1 - \Omega_a}\left( \frac{1}{2}g_{{\sss XC},ijk}^{\rm adia}a_{a,jn} a_{a,kp} + a_{{\sss S}2,ij}^{-1} \frac{\nu_1^2}{\Omega_a^2} \langle f\Hxc(\omega_2)\rangle_a A_{{\sss S}12, jpn}\right)
%+\frac{1}{\omega_2 - \Omega_a}\left( \frac{1}{2}g_{{\sss XC},ijk}^{\rm adia}a_{a,jn} a_{a,kp} + a_{{\sss S}2,ij}^{-1} \frac{\nu_1^2}{\Omega_a^2} \langle f\Hxc(\omega_2)\rangle_a A_{{\sss S}12, jpn}\right)\right)\right]
\label{eq:s_app1}
\eea
}}

If we were then to extract the transition density between states $a$ and $c$ (Eq. 11 of the main text), we obtain
\ben
n_{ca}^{\rm App-1}(\br) =  \left(\frac{\nu_1}{\Omega_a}\right)^{3/2}\left(1 - 2\frac{\langle f\Hxc \rangle_1}{\Omega_a}\right)\alpha_c n_{{\sss S},13}(\br) - \frac{n_{0a}(\br)}{\Omega_a}\int n_{0c}(\br_1)g\xc^{\rm adia}(r_1,r_2,r_3) n_{0a}(\br_2)n_{0a}(\br_3)d\br_1 d\br_2 d\br_3
\label{eq:s_approx-nac1}
\een
where we have defined $\alpha_c^2$ as an $\br$-independent approximation to $a_{c,ij}a^{-1}_{{\sss S}3,jk}$:
\ben
a_c \approx \alpha_c^2 a_{{\sss S} 3}
\label{eq:s_alphac}
\een
Physically, $\alpha_c$ reflects the ratio of the true transition density to the KS  one. Due to the oscillator-strength sum-rule applied within the separated levels assumption, $\alpha_c^2 \approx \nu_3/\Omega_c$ when the interacting state is predominantly a single-excitation. When there is mixing with a KS double-excitation however, $\alpha_c^2$ decreases, reducing to zero in the limit of a pure double-excitation. 
This approximation (Eq.~(\ref{eq:s_approx-nac1})) however appears not to include double-excitation contributions to the transition density. The first term corrects the single-excitation component, while it is unclear whether the second term, which tends to be much smaller than the first, captures true double-excitation character: an adiabatic QR kernel yields a response that has poles at sums of LR-corrected single excitations without any mixing with double-excitations but even these poles are missing when only forward transitions are kept~\cite{EGCM11}. Our second limiting case therefore focusses on the double-excitation contribution. 

(ii) {\underline{\it Second limiting case:}} Here we consider the case when state $c$ is close to a pure double-excitation of the KS system, denoted $d$. There is no single KS excitation in the vicinity, so no pole in the KS LR function nearby.  This means that the KS residue $A_{{\sss S}1d, ijk} = n_{01}(\br_i)n_{1d}(\br_j)n_{d0}(\br_k)$ is strictly zero because $n_{d0}(\br) \equiv 0$ since a one-body operator cannot connect determinants different by two orbitals.  The interacting residue is small but non-zero because the corresponding term $n_{c0}$ involves the state $c$ which has small contributions from single-excitations to the KS double excitation that dominates the interacting state in this limit. In this limit the exact interacting $\chi^{(2)}$ has the form
\ben
\chi^{(2)}_{mnp}(\omega_1,\omega_2) = \frac{1}{2(\omega_1 +\omega_2 - \Omega_c)}\left(\frac{A_{ca,mnp}}{\omega_2 - \Omega_a} + \frac{A_{ca,mpn}}{\omega_1 - \Omega_a} \right), \;\;\; A_{ca} = n_{0c}n_{ca}n_{a0}
\een
so that equating Eq.~(\ref{eq:s_tddftchi2_1}) to this, gives Eq.~(9) of the main paper. We now approximate the part of the residue  $A_{ca}$ that is not accessible from LR, $n_{ca}$ simply by the KS transition-density $n_{d1}$, yielding 
\ben
a_{c,mi}g^{\rm App-2}_{{\sss XC},ijk} a_{a,jn} a_{a,kp} = \frac{n_{0c}(\br_m)}{2}\left[\left(\omega_1 - \Omega_a\right)n_{d1}(\br_n)n_{0a}(\br_p) + \left(\omega_2 - \Omega_a\right)n_{d1}(\br_p)n_{0a}(\br_n) \right]
\label{eq:s_app-2}
\een

Our complete approximation interpolates between limits (1) and (2) through
\ben
g\xc^{\rm App} = g\xc^{\rm App-1} + \sqrt{1- a_c a^{-1}_{{\sss S}3} }g\xc^{\rm App-2}
\een
as in Eq.~(10) of the main text,  with the rationale that $1 - a_c a^{-1}_{{\sss S}3}$ represents the fraction of the true oscillator strength arising from the double-excitation component of the state.

\bibliography{./main.bib}

\begin{thebibliography}{45}%
\makeatletter
\providecommand \@ifxundefined [1]{%
 \@ifx{#1\undefined}
}%
\providecommand \@ifnum [1]{%
 \ifnum #1\expandafter \@firstoftwo
 \else \expandafter \@secondoftwo
 \fi
}%
\providecommand \@ifx [1]{%
 \ifx #1\expandafter \@firstoftwo
 \else \expandafter \@secondoftwo
 \fi
}%
\providecommand \natexlab [1]{#1}%
\providecommand \enquote  [1]{``#1''}%
\providecommand \bibnamefont  [1]{#1}%
\providecommand \bibfnamefont [1]{#1}%
\providecommand \citenamefont [1]{#1}%
\providecommand \href@noop [0]{\@secondoftwo}%
\providecommand \href [0]{\begingroup \@sanitize@url \@href}%
\providecommand \@href[1]{\@@startlink{#1}\@@href}%
\providecommand \@@href[1]{\endgroup#1\@@endlink}%
\providecommand \@sanitize@url [0]{\catcode `\\12\catcode `\$12\catcode
  `\&12\catcode `\#12\catcode `\^12\catcode `\_12\catcode `\%12\relax}%
\providecommand \@@startlink[1]{}%
\providecommand \@@endlink[0]{}%
\providecommand \url  [0]{\begingroup\@sanitize@url \@url }%
\providecommand \@url [1]{\endgroup\@href {#1}{\urlprefix }}%
\providecommand \urlprefix  [0]{URL }%
\providecommand \Eprint [0]{\href }%
\providecommand \doibase [0]{http://dx.doi.org/}%
\providecommand \selectlanguage [0]{\@gobble}%
\providecommand \bibinfo  [0]{\@secondoftwo}%
\providecommand \bibfield  [0]{\@secondoftwo}%
\providecommand \translation [1]{[#1]}%
\providecommand \BibitemOpen [0]{}%
\providecommand \bibitemStop [0]{}%
\providecommand \bibitemNoStop [0]{.\EOS\space}%
\providecommand \EOS [0]{\spacefactor3000\relax}%
\providecommand \BibitemShut  [1]{\csname bibitem#1\endcsname}%
\let\auto@bib@innerbib\@empty
%</preamble>
\bibitem [{\citenamefont {Papadopoulos}\ \emph {et~al.}(2006)\citenamefont
  {Papadopoulos}, \citenamefont {Sadlej},\ and\ \citenamefont
  {Leszczynski}}]{PapadopoulosBook}%
  \BibitemOpen
  \bibinfo {editor} {\bibfnamefont {M.~G.}\ \bibnamefont {Papadopoulos}},
  \bibinfo {editor} {\bibfnamefont {A.~J.}\ \bibnamefont {Sadlej}}, \ and\
  \bibinfo {editor} {\bibfnamefont {J.}~\bibnamefont {Leszczynski}},\ eds.,\
  \href {\doibase 10.1007/1-4020-4850-5} {\emph {\bibinfo {title} {Non-Linear
  Optical Properties of Matter}}}\ (\bibinfo  {publisher} {Springer
  Netherlands},\ \bibinfo {year} {2006})\BibitemShut {NoStop}%
\bibitem [{\citenamefont {Mukamel}(1999)}]{MukamelBook}%
  \BibitemOpen
  \bibfield  {author} {\bibinfo {author} {\bibfnamefont {S.}~\bibnamefont
  {Mukamel}},\ }\href@noop {} {\emph {\bibinfo {title} {Principles of
  Non-Linear Optical Spectroscopy}}}\ (\bibinfo  {publisher} {Oxford University
  Press},\ \bibinfo {year} {1999})\BibitemShut {NoStop}%
\bibitem [{\citenamefont {Parker}\ \emph {et~al.}(2016)\citenamefont {Parker},
  \citenamefont {Roy},\ and\ \citenamefont {Furche}}]{PRF16}%
  \BibitemOpen
  \bibfield  {author} {\bibinfo {author} {\bibfnamefont {S.~M.}\ \bibnamefont
  {Parker}}, \bibinfo {author} {\bibfnamefont {S.}~\bibnamefont {Roy}}, \ and\
  \bibinfo {author} {\bibfnamefont {F.}~\bibnamefont {Furche}},\ }\href@noop {}
  {\bibfield  {journal} {\bibinfo  {journal} {Journal of chemical physics}\
  }\textbf {\bibinfo {volume} {145}},\ \bibinfo {pages} {134105} (\bibinfo
  {year} {2016})}\BibitemShut {NoStop}%
\bibitem [{\citenamefont {Dalgaard}(1982)}]{D82}%
  \BibitemOpen
  \bibfield  {author} {\bibinfo {author} {\bibfnamefont {E.}~\bibnamefont
  {Dalgaard}},\ }\href {\doibase 10.1103/PhysRevA.26.42} {\bibfield  {journal}
  {\bibinfo  {journal} {Phys. Rev. A}\ }\textbf {\bibinfo {volume} {26}},\
  \bibinfo {pages} {42} (\bibinfo {year} {1982})}\BibitemShut {NoStop}%
\bibitem [{\citenamefont {Li}\ and\ \citenamefont {Liu}(2014)}]{LL14}%
  \BibitemOpen
  \bibfield  {author} {\bibinfo {author} {\bibfnamefont {Z.}~\bibnamefont
  {Li}}\ and\ \bibinfo {author} {\bibfnamefont {W.}~\bibnamefont {Liu}},\
  }\href@noop {} {\bibfield  {journal} {\bibinfo  {journal} {The Journal of
  Chemical Physics}\ }\textbf {\bibinfo {volume} {141}},\ \bibinfo {pages}
  {014110} (\bibinfo {year} {2014})}\BibitemShut {NoStop}%
\bibitem [{\citenamefont {Ou}\ \emph {et~al.}(2015{\natexlab{a}})\citenamefont
  {Ou}, \citenamefont {Bellchambers}, \citenamefont {Furche},\ and\
  \citenamefont {Subotnik}}]{OBFS15}%
  \BibitemOpen
  \bibfield  {author} {\bibinfo {author} {\bibfnamefont {Q.}~\bibnamefont
  {Ou}}, \bibinfo {author} {\bibfnamefont {G.~D.}\ \bibnamefont
  {Bellchambers}}, \bibinfo {author} {\bibfnamefont {F.}~\bibnamefont
  {Furche}}, \ and\ \bibinfo {author} {\bibfnamefont {J.~E.}\ \bibnamefont
  {Subotnik}},\ }\href@noop {} {\bibfield  {journal} {\bibinfo  {journal} {The
  Journal of chemical physics}\ }\textbf {\bibinfo {volume} {142}},\ \bibinfo
  {pages} {064114} (\bibinfo {year} {2015}{\natexlab{a}})}\BibitemShut
  {NoStop}%
\bibitem [{\citenamefont {Zhang}\ and\ \citenamefont {Herbert}(2015)}]{ZH15}%
  \BibitemOpen
  \bibfield  {author} {\bibinfo {author} {\bibfnamefont {X.}~\bibnamefont
  {Zhang}}\ and\ \bibinfo {author} {\bibfnamefont {J.~M.}\ \bibnamefont
  {Herbert}},\ }\href@noop {} {\bibfield  {journal} {\bibinfo  {journal} {The
  Journal of Chemical Physics}\ }\textbf {\bibinfo {volume} {142}},\ \bibinfo
  {pages} {064109} (\bibinfo {year} {2015})}\BibitemShut {NoStop}%
\bibitem [{\citenamefont {Runge}\ and\ \citenamefont {Gross}(1984)}]{RG84}%
  \BibitemOpen
  \bibfield  {author} {\bibinfo {author} {\bibfnamefont {E.}~\bibnamefont
  {Runge}}\ and\ \bibinfo {author} {\bibfnamefont {E.~K.~U.}\ \bibnamefont
  {Gross}},\ }\href@noop {} {\bibfield  {journal} {\bibinfo  {journal} {Phys.
  Rev. Lett.}\ }\textbf {\bibinfo {volume} {52}},\ \bibinfo {pages} {997}
  (\bibinfo {year} {1984})}\BibitemShut {NoStop}%
\bibitem [{\citenamefont {Ullrich}(2011)}]{Carstenbook}%
  \BibitemOpen
  \bibfield  {author} {\bibinfo {author} {\bibfnamefont {C.~A.}\ \bibnamefont
  {Ullrich}},\ }\href@noop {} {\emph {\bibinfo {title} {Time-dependent
  density-functional theory: concepts and applications}}}\ (\bibinfo
  {publisher} {Oxford University Press},\ \bibinfo {year} {2011})\BibitemShut
  {NoStop}%
\bibitem [{\citenamefont {Marques}\ \emph {et~al.}(2012)\citenamefont
  {Marques}, \citenamefont {Maitra}, \citenamefont {Nogueira}, \citenamefont
  {Gross},\ and\ \citenamefont {Rubio}}]{TDDFTbook2012}%
  \BibitemOpen
  \bibinfo {editor} {\bibfnamefont {M.~A.}\ \bibnamefont {Marques}}, \bibinfo
  {editor} {\bibfnamefont {N.~T.}\ \bibnamefont {Maitra}}, \bibinfo {editor}
  {\bibfnamefont {F.~M.}\ \bibnamefont {Nogueira}}, \bibinfo {editor}
  {\bibfnamefont {E.~K.}\ \bibnamefont {Gross}}, \ and\ \bibinfo {editor}
  {\bibfnamefont {A.}~\bibnamefont {Rubio}},\ eds.,\ \href@noop {} {\emph
  {\bibinfo {title} {Fundamentals of time-dependent density functional
  theory}}},\ Vol.\ \bibinfo {volume} {837}\ (\bibinfo  {publisher}
  {Springer},\ \bibinfo {year} {2012})\BibitemShut {NoStop}%
\bibitem [{\citenamefont {Maitra}(2016)}]{M16}%
  \BibitemOpen
  \bibfield  {author} {\bibinfo {author} {\bibfnamefont {N.~T.}\ \bibnamefont
  {Maitra}},\ }\href@noop {} {\bibfield  {journal} {\bibinfo  {journal} {The
  Journal of Chemical Physics}\ }\textbf {\bibinfo {volume} {144}},\ \bibinfo
  {pages} {220901} (\bibinfo {year} {2016})}\BibitemShut {NoStop}%
\bibitem [{\citenamefont {Gross}\ \emph {et~al.}(1996)\citenamefont {Gross},
  \citenamefont {Dobson},\ and\ \citenamefont {Petersilka}}]{GDP96}%
  \BibitemOpen
  \bibfield  {author} {\bibinfo {author} {\bibfnamefont {E.~K.~U.}\
  \bibnamefont {Gross}}, \bibinfo {author} {\bibfnamefont {J.~F.}\ \bibnamefont
  {Dobson}}, \ and\ \bibinfo {author} {\bibfnamefont {M.}~\bibnamefont
  {Petersilka}},\ }in\ \href@noop {} {\emph {\bibinfo {booktitle} {Density
  Functional Theory II: Relativistic and Time Dependent Extensions}}},\
  \bibinfo {editor} {edited by\ \bibinfo {editor} {\bibfnamefont {R.~F.}\
  \bibnamefont {Nalewajski}}}\ (\bibinfo  {publisher} {Springer Berlin
  Heidelberg},\ \bibinfo {address} {Berlin, Heidelberg},\ \bibinfo {year}
  {1996})\ pp.\ \bibinfo {pages} {81--172}\BibitemShut {NoStop}%
\bibitem [{\citenamefont {Petersilka}\ \emph {et~al.}(1996)\citenamefont
  {Petersilka}, \citenamefont {Gossmann},\ and\ \citenamefont {Gross}}]{PGG96}%
  \BibitemOpen
  \bibfield  {author} {\bibinfo {author} {\bibfnamefont {M.}~\bibnamefont
  {Petersilka}}, \bibinfo {author} {\bibfnamefont {U.~J.}\ \bibnamefont
  {Gossmann}}, \ and\ \bibinfo {author} {\bibfnamefont {E.~K.~U.}\ \bibnamefont
  {Gross}},\ }\href@noop {} {\bibfield  {journal} {\bibinfo  {journal} {Phys.
  Rev. Lett.}\ }\textbf {\bibinfo {volume} {76}},\ \bibinfo {pages} {1212}
  (\bibinfo {year} {1996})}\BibitemShut {NoStop}%
\bibitem [{\citenamefont {Casida}(1995)}]{C95}%
  \BibitemOpen
  \bibfield  {author} {\bibinfo {author} {\bibfnamefont {M.}~\bibnamefont
  {Casida}},\ }in\ \href@noop {} {\emph {\bibinfo {booktitle} {Recent Advances
  in Density Functional Methods, Part I}}},\ \bibinfo {editor} {edited by\
  \bibinfo {editor} {\bibfnamefont {D.}~\bibnamefont {Chong}}}\ (\bibinfo
  {publisher} {World Scientific, Singapore},\ \bibinfo {year}
  {1995})\BibitemShut {NoStop}%
\bibitem [{\citenamefont {Maitra}\ \emph {et~al.}(2004)\citenamefont {Maitra},
  \citenamefont {Zhang}, \citenamefont {Cave},\ and\ \citenamefont
  {Burke}}]{MZCB04}%
  \BibitemOpen
  \bibfield  {author} {\bibinfo {author} {\bibfnamefont {N.~T.}\ \bibnamefont
  {Maitra}}, \bibinfo {author} {\bibfnamefont {F.}~\bibnamefont {Zhang}},
  \bibinfo {author} {\bibfnamefont {R.~J.}\ \bibnamefont {Cave}}, \ and\
  \bibinfo {author} {\bibfnamefont {K.}~\bibnamefont {Burke}},\ }\href@noop {}
  {\bibfield  {journal} {\bibinfo  {journal} {The Journal of Chemical Physics}\
  }\textbf {\bibinfo {volume} {120}},\ \bibinfo {pages} {5932} (\bibinfo {year}
  {2004})}\BibitemShut {NoStop}%
\bibitem [{\citenamefont {Maitra}(2022)}]{M22}%
  \BibitemOpen
  \bibfield  {author} {\bibinfo {author} {\bibfnamefont {N.~T.}\ \bibnamefont
  {Maitra}},\ }\href {\doibase 10.1146/annurev-physchem-082720-124933}
  {\bibfield  {journal} {\bibinfo  {journal} {Annual Review of Physical
  Chemistry}\ }\textbf {\bibinfo {volume} {73}},\ \bibinfo {pages} {117}
  (\bibinfo {year} {2022})}\BibitemShut {NoStop}%
\bibitem [{\citenamefont {Wehrum}\ and\ \citenamefont
  {Hermeking}(1974)}]{WH74}%
  \BibitemOpen
  \bibfield  {author} {\bibinfo {author} {\bibfnamefont {R.~P.}\ \bibnamefont
  {Wehrum}}\ and\ \bibinfo {author} {\bibfnamefont {H.}~\bibnamefont
  {Hermeking}},\ }\href {\doibase 10.1088/0022-3719/7/6/003} {\bibfield
  {journal} {\bibinfo  {journal} {Journal of Physics C: Solid State Physics}\
  }\textbf {\bibinfo {volume} {7}},\ \bibinfo {pages} {L107} (\bibinfo {year}
  {1974})}\BibitemShut {NoStop}%
\bibitem [{\citenamefont {Senatore}\ and\ \citenamefont
  {Subbaswamy}(1987)}]{SS87}%
  \BibitemOpen
  \bibfield  {author} {\bibinfo {author} {\bibfnamefont {G.}~\bibnamefont
  {Senatore}}\ and\ \bibinfo {author} {\bibfnamefont {K.}~\bibnamefont
  {Subbaswamy}},\ }\href@noop {} {\bibfield  {journal} {\bibinfo  {journal}
  {Physical Review A}\ }\textbf {\bibinfo {volume} {35}},\ \bibinfo {pages}
  {2440} (\bibinfo {year} {1987})}\BibitemShut {NoStop}%
\bibitem [{\citenamefont {Parker}\ and\ \citenamefont {Furche}(2018)}]{PF17}%
  \BibitemOpen
  \bibfield  {author} {\bibinfo {author} {\bibfnamefont {S.~M.}\ \bibnamefont
  {Parker}}\ and\ \bibinfo {author} {\bibfnamefont {F.}~\bibnamefont
  {Furche}},\ }\enquote {\bibinfo {title} {Response theory and molecular
  properties},}\ in\ \href {\doibase 10.1007/978-981-10-5651-2_4} {\emph
  {\bibinfo {booktitle} {Frontiers of Quantum Chemistry}}},\ \bibinfo {editor}
  {edited by\ \bibinfo {editor} {\bibfnamefont {M.~J.}\ \bibnamefont
  {W{\'o}jcik}}, \bibinfo {editor} {\bibfnamefont {H.}~\bibnamefont
  {Nakatsuji}}, \bibinfo {editor} {\bibfnamefont {B.}~\bibnamefont {Kirtman}},
  \ and\ \bibinfo {editor} {\bibfnamefont {Y.}~\bibnamefont {Ozaki}}}\
  (\bibinfo  {publisher} {Springer Singapore},\ \bibinfo {address}
  {Singapore},\ \bibinfo {year} {2018})\ pp.\ \bibinfo {pages}
  {69--86}\BibitemShut {NoStop}%
\bibitem [{\citenamefont {Sa{\l}ek}\ \emph {et~al.}(2002)\citenamefont
  {Sa{\l}ek}, \citenamefont {Vahtras}, \citenamefont {Helgaker},\ and\
  \citenamefont {{\AA}gren}}]{SVHA02}%
  \BibitemOpen
  \bibfield  {author} {\bibinfo {author} {\bibfnamefont {P.}~\bibnamefont
  {Sa{\l}ek}}, \bibinfo {author} {\bibfnamefont {O.}~\bibnamefont {Vahtras}},
  \bibinfo {author} {\bibfnamefont {T.}~\bibnamefont {Helgaker}}, \ and\
  \bibinfo {author} {\bibfnamefont {H.}~\bibnamefont {{\AA}gren}},\ }\href@noop
  {} {\bibfield  {journal} {\bibinfo  {journal} {The Journal of chemical
  physics}\ }\textbf {\bibinfo {volume} {117}},\ \bibinfo {pages} {9630}
  (\bibinfo {year} {2002})}\BibitemShut {NoStop}%
\bibitem [{\citenamefont {Aidas}\ \emph {et~al.}(2013)\citenamefont {Aidas},
  \citenamefont {Angeli}, \citenamefont {Bak}, \citenamefont {Bakken},
  \citenamefont {Bast}, \citenamefont {Boman}, \citenamefont {Christiansen},
  \citenamefont {Cimiraglia}, \citenamefont {Coriani}, \citenamefont {Dahle},
  \citenamefont {Dalskov}, \citenamefont {Ekström}, \citenamefont
  {Enevoldsen}, \citenamefont {Eriksen}, \citenamefont {Ettenhuber},
  \citenamefont {Fern{\'{a}}ndez}, \citenamefont {Ferrighi}, \citenamefont
  {Fliegl}, \citenamefont {Frediani}, \citenamefont {Hald}, \citenamefont
  {Halkier}, \citenamefont {Hättig}, \citenamefont {Heiberg}, \citenamefont
  {Helgaker}, \citenamefont {Hennum}, \citenamefont {Hettema}, \citenamefont
  {Hjertenaes}, \citenamefont {H{\o}st}, \citenamefont {H{\o}yvik},
  \citenamefont {Iozzi}, \citenamefont {Jans{\'{\i}}k}, \citenamefont {Jensen},
  \citenamefont {Jonsson}, \citenamefont {J{\o}rgensen}, \citenamefont
  {Kauczor}, \citenamefont {Kirpekar}, \citenamefont {Kjaergaard},
  \citenamefont {Klopper}, \citenamefont {Knecht}, \citenamefont {Kobayashi},
  \citenamefont {Koch}, \citenamefont {Kongsted}, \citenamefont {Krapp},
  \citenamefont {Kristensen}, \citenamefont {Ligabue}, \citenamefont {Lutnaes},
  \citenamefont {Melo}, \citenamefont {Mikkelsen}, \citenamefont {Myhre},
  \citenamefont {Neiss}, \citenamefont {Nielsen}, \citenamefont {Norman},
  \citenamefont {Olsen}, \citenamefont {Olsen}, \citenamefont {Osted},
  \citenamefont {Packer}, \citenamefont {Pawlowski}, \citenamefont {Pedersen},
  \citenamefont {Provasi}, \citenamefont {Reine}, \citenamefont {Rinkevicius},
  \citenamefont {Ruden}, \citenamefont {Ruud}, \citenamefont {Rybkin},
  \citenamefont {Sa{\l}ek}, \citenamefont {Samson}, \citenamefont
  {de~Mer{\'{a}}s}, \citenamefont {Saue}, \citenamefont {Sauer}, \citenamefont
  {Schimmelpfennig}, \citenamefont {Sneskov}, \citenamefont {Steindal},
  \citenamefont {Sylvester-Hvid}, \citenamefont {Taylor}, \citenamefont
  {Teale}, \citenamefont {Tellgren}, \citenamefont {Tew}, \citenamefont
  {Thorvaldsen}, \citenamefont {Th{\o}gersen}, \citenamefont {Vahtras},
  \citenamefont {Watson}, \citenamefont {Wilson}, \citenamefont {Ziolkowski},\
  and\ \citenamefont {{\AA}gren}}]{dalton}%
  \BibitemOpen
  \bibfield  {author} {\bibinfo {author} {\bibfnamefont {K.}~\bibnamefont
  {Aidas}}, \bibinfo {author} {\bibfnamefont {C.}~\bibnamefont {Angeli}},
  \bibinfo {author} {\bibfnamefont {K.~L.}\ \bibnamefont {Bak}}, \bibinfo
  {author} {\bibfnamefont {V.}~\bibnamefont {Bakken}}, \bibinfo {author}
  {\bibfnamefont {R.}~\bibnamefont {Bast}}, \bibinfo {author} {\bibfnamefont
  {L.}~\bibnamefont {Boman}}, \bibinfo {author} {\bibfnamefont
  {O.}~\bibnamefont {Christiansen}}, \bibinfo {author} {\bibfnamefont
  {R.}~\bibnamefont {Cimiraglia}}, \bibinfo {author} {\bibfnamefont
  {S.}~\bibnamefont {Coriani}}, \bibinfo {author} {\bibfnamefont
  {P.}~\bibnamefont {Dahle}}, \bibinfo {author} {\bibfnamefont {E.~K.}\
  \bibnamefont {Dalskov}}, \bibinfo {author} {\bibfnamefont {U.}~\bibnamefont
  {Ekström}}, \bibinfo {author} {\bibfnamefont {T.}~\bibnamefont
  {Enevoldsen}}, \bibinfo {author} {\bibfnamefont {J.~J.}\ \bibnamefont
  {Eriksen}}, \bibinfo {author} {\bibfnamefont {P.}~\bibnamefont {Ettenhuber}},
  \bibinfo {author} {\bibfnamefont {B.}~\bibnamefont {Fern{\'{a}}ndez}},
  \bibinfo {author} {\bibfnamefont {L.}~\bibnamefont {Ferrighi}}, \bibinfo
  {author} {\bibfnamefont {H.}~\bibnamefont {Fliegl}}, \bibinfo {author}
  {\bibfnamefont {L.}~\bibnamefont {Frediani}}, \bibinfo {author}
  {\bibfnamefont {K.}~\bibnamefont {Hald}}, \bibinfo {author} {\bibfnamefont
  {A.}~\bibnamefont {Halkier}}, \bibinfo {author} {\bibfnamefont
  {C.}~\bibnamefont {Hättig}}, \bibinfo {author} {\bibfnamefont
  {H.}~\bibnamefont {Heiberg}}, \bibinfo {author} {\bibfnamefont
  {T.}~\bibnamefont {Helgaker}}, \bibinfo {author} {\bibfnamefont {A.~C.}\
  \bibnamefont {Hennum}}, \bibinfo {author} {\bibfnamefont {H.}~\bibnamefont
  {Hettema}}, \bibinfo {author} {\bibfnamefont {E.}~\bibnamefont {Hjertenaes}},
  \bibinfo {author} {\bibfnamefont {S.}~\bibnamefont {H{\o}st}}, \bibinfo
  {author} {\bibfnamefont {I.-M.}\ \bibnamefont {H{\o}yvik}}, \bibinfo {author}
  {\bibfnamefont {M.~F.}\ \bibnamefont {Iozzi}}, \bibinfo {author}
  {\bibfnamefont {B.}~\bibnamefont {Jans{\'{\i}}k}}, \bibinfo {author}
  {\bibfnamefont {H.~J.~A.}\ \bibnamefont {Jensen}}, \bibinfo {author}
  {\bibfnamefont {D.}~\bibnamefont {Jonsson}}, \bibinfo {author} {\bibfnamefont
  {P.}~\bibnamefont {J{\o}rgensen}}, \bibinfo {author} {\bibfnamefont
  {J.}~\bibnamefont {Kauczor}}, \bibinfo {author} {\bibfnamefont
  {S.}~\bibnamefont {Kirpekar}}, \bibinfo {author} {\bibfnamefont
  {T.}~\bibnamefont {Kjaergaard}}, \bibinfo {author} {\bibfnamefont
  {W.}~\bibnamefont {Klopper}}, \bibinfo {author} {\bibfnamefont
  {S.}~\bibnamefont {Knecht}}, \bibinfo {author} {\bibfnamefont
  {R.}~\bibnamefont {Kobayashi}}, \bibinfo {author} {\bibfnamefont
  {H.}~\bibnamefont {Koch}}, \bibinfo {author} {\bibfnamefont {J.}~\bibnamefont
  {Kongsted}}, \bibinfo {author} {\bibfnamefont {A.}~\bibnamefont {Krapp}},
  \bibinfo {author} {\bibfnamefont {K.}~\bibnamefont {Kristensen}}, \bibinfo
  {author} {\bibfnamefont {A.}~\bibnamefont {Ligabue}}, \bibinfo {author}
  {\bibfnamefont {O.~B.}\ \bibnamefont {Lutnaes}}, \bibinfo {author}
  {\bibfnamefont {J.~I.}\ \bibnamefont {Melo}}, \bibinfo {author}
  {\bibfnamefont {K.~V.}\ \bibnamefont {Mikkelsen}}, \bibinfo {author}
  {\bibfnamefont {R.~H.}\ \bibnamefont {Myhre}}, \bibinfo {author}
  {\bibfnamefont {C.}~\bibnamefont {Neiss}}, \bibinfo {author} {\bibfnamefont
  {C.~B.}\ \bibnamefont {Nielsen}}, \bibinfo {author} {\bibfnamefont
  {P.}~\bibnamefont {Norman}}, \bibinfo {author} {\bibfnamefont
  {J.}~\bibnamefont {Olsen}}, \bibinfo {author} {\bibfnamefont {J.~M.~H.}\
  \bibnamefont {Olsen}}, \bibinfo {author} {\bibfnamefont {A.}~\bibnamefont
  {Osted}}, \bibinfo {author} {\bibfnamefont {M.~J.}\ \bibnamefont {Packer}},
  \bibinfo {author} {\bibfnamefont {F.}~\bibnamefont {Pawlowski}}, \bibinfo
  {author} {\bibfnamefont {T.~B.}\ \bibnamefont {Pedersen}}, \bibinfo {author}
  {\bibfnamefont {P.~F.}\ \bibnamefont {Provasi}}, \bibinfo {author}
  {\bibfnamefont {S.}~\bibnamefont {Reine}}, \bibinfo {author} {\bibfnamefont
  {Z.}~\bibnamefont {Rinkevicius}}, \bibinfo {author} {\bibfnamefont {T.~A.}\
  \bibnamefont {Ruden}}, \bibinfo {author} {\bibfnamefont {K.}~\bibnamefont
  {Ruud}}, \bibinfo {author} {\bibfnamefont {V.~V.}\ \bibnamefont {Rybkin}},
  \bibinfo {author} {\bibfnamefont {P.}~\bibnamefont {Sa{\l}ek}}, \bibinfo
  {author} {\bibfnamefont {C.~C.~M.}\ \bibnamefont {Samson}}, \bibinfo {author}
  {\bibfnamefont {A.~S.}\ \bibnamefont {de~Mer{\'{a}}s}}, \bibinfo {author}
  {\bibfnamefont {T.}~\bibnamefont {Saue}}, \bibinfo {author} {\bibfnamefont
  {S.~P.~A.}\ \bibnamefont {Sauer}}, \bibinfo {author} {\bibfnamefont
  {B.}~\bibnamefont {Schimmelpfennig}}, \bibinfo {author} {\bibfnamefont
  {K.}~\bibnamefont {Sneskov}}, \bibinfo {author} {\bibfnamefont {A.~H.}\
  \bibnamefont {Steindal}}, \bibinfo {author} {\bibfnamefont {K.~O.}\
  \bibnamefont {Sylvester-Hvid}}, \bibinfo {author} {\bibfnamefont {P.~R.}\
  \bibnamefont {Taylor}}, \bibinfo {author} {\bibfnamefont {A.~M.}\
  \bibnamefont {Teale}}, \bibinfo {author} {\bibfnamefont {E.~I.}\ \bibnamefont
  {Tellgren}}, \bibinfo {author} {\bibfnamefont {D.~P.}\ \bibnamefont {Tew}},
  \bibinfo {author} {\bibfnamefont {A.~J.}\ \bibnamefont {Thorvaldsen}},
  \bibinfo {author} {\bibfnamefont {L.}~\bibnamefont {Th{\o}gersen}}, \bibinfo
  {author} {\bibfnamefont {O.}~\bibnamefont {Vahtras}}, \bibinfo {author}
  {\bibfnamefont {M.~A.}\ \bibnamefont {Watson}}, \bibinfo {author}
  {\bibfnamefont {D.~J.~D.}\ \bibnamefont {Wilson}}, \bibinfo {author}
  {\bibfnamefont {M.}~\bibnamefont {Ziolkowski}}, \ and\ \bibinfo {author}
  {\bibfnamefont {H.}~\bibnamefont {{\AA}gren}},\ }\href {\doibase
  10.1002/wcms.1172} {\bibfield  {journal} {\bibinfo  {journal} {Wiley
  Interdisciplinary Reviews: Computational Molecular Science}\ }\textbf
  {\bibinfo {volume} {4}},\ \bibinfo {pages} {269} (\bibinfo {year}
  {2013})}\BibitemShut {NoStop}%
\bibitem [{\citenamefont {Balasubramani}\ \emph {et~al.}(2020)\citenamefont
  {Balasubramani}, \citenamefont {Chen}, \citenamefont {Coriani}, \citenamefont
  {Diedenhofen}, \citenamefont {Frank}, \citenamefont {Franzke}, \citenamefont
  {Furche}, \citenamefont {Grotjahn}, \citenamefont {Harding}, \citenamefont
  {H{\"a}ttig}, \citenamefont {Hellweg}, \citenamefont {Helmich-Paris},
  \citenamefont {Holzer}, \citenamefont {Huniar}, \citenamefont {Kaupp},
  \citenamefont {Marefat~Khah}, \citenamefont {Karbalaei~Khani}, \citenamefont
  {M{\"u}ller}, \citenamefont {Mack}, \citenamefont {Nguyen}, \citenamefont
  {Parker}, \citenamefont {Perlt}, \citenamefont {Rappoport}, \citenamefont
  {Reiter}, \citenamefont {Roy}, \citenamefont {R{\"u}ckert}, \citenamefont
  {Schmitz}, \citenamefont {Sierka}, \citenamefont {Tapavicza}, \citenamefont
  {Tew}, \citenamefont {van W{\"u}llen}, \citenamefont {Voora}, \citenamefont
  {Weigend}, \citenamefont {Wody{\'n}ski},\ and\ \citenamefont
  {Yu}}]{turbomole2}%
  \BibitemOpen
  \bibfield  {author} {\bibinfo {author} {\bibfnamefont {S.~G.}\ \bibnamefont
  {Balasubramani}}, \bibinfo {author} {\bibfnamefont {G.~P.}\ \bibnamefont
  {Chen}}, \bibinfo {author} {\bibfnamefont {S.}~\bibnamefont {Coriani}},
  \bibinfo {author} {\bibfnamefont {M.}~\bibnamefont {Diedenhofen}}, \bibinfo
  {author} {\bibfnamefont {M.~S.}\ \bibnamefont {Frank}}, \bibinfo {author}
  {\bibfnamefont {Y.~J.}\ \bibnamefont {Franzke}}, \bibinfo {author}
  {\bibfnamefont {F.}~\bibnamefont {Furche}}, \bibinfo {author} {\bibfnamefont
  {R.}~\bibnamefont {Grotjahn}}, \bibinfo {author} {\bibfnamefont {M.~E.}\
  \bibnamefont {Harding}}, \bibinfo {author} {\bibfnamefont {C.}~\bibnamefont
  {H{\"a}ttig}}, \bibinfo {author} {\bibfnamefont {A.}~\bibnamefont {Hellweg}},
  \bibinfo {author} {\bibfnamefont {B.}~\bibnamefont {Helmich-Paris}}, \bibinfo
  {author} {\bibfnamefont {C.}~\bibnamefont {Holzer}}, \bibinfo {author}
  {\bibfnamefont {U.}~\bibnamefont {Huniar}}, \bibinfo {author} {\bibfnamefont
  {M.}~\bibnamefont {Kaupp}}, \bibinfo {author} {\bibfnamefont
  {A.}~\bibnamefont {Marefat~Khah}}, \bibinfo {author} {\bibfnamefont
  {S.}~\bibnamefont {Karbalaei~Khani}}, \bibinfo {author} {\bibfnamefont
  {T.}~\bibnamefont {M{\"u}ller}}, \bibinfo {author} {\bibfnamefont
  {F.}~\bibnamefont {Mack}}, \bibinfo {author} {\bibfnamefont {B.~D.}\
  \bibnamefont {Nguyen}}, \bibinfo {author} {\bibfnamefont {S.~M.}\
  \bibnamefont {Parker}}, \bibinfo {author} {\bibfnamefont {E.}~\bibnamefont
  {Perlt}}, \bibinfo {author} {\bibfnamefont {D.}~\bibnamefont {Rappoport}},
  \bibinfo {author} {\bibfnamefont {K.}~\bibnamefont {Reiter}}, \bibinfo
  {author} {\bibfnamefont {S.}~\bibnamefont {Roy}}, \bibinfo {author}
  {\bibfnamefont {M.}~\bibnamefont {R{\"u}ckert}}, \bibinfo {author}
  {\bibfnamefont {G.}~\bibnamefont {Schmitz}}, \bibinfo {author} {\bibfnamefont
  {M.}~\bibnamefont {Sierka}}, \bibinfo {author} {\bibfnamefont
  {E.}~\bibnamefont {Tapavicza}}, \bibinfo {author} {\bibfnamefont {D.~P.}\
  \bibnamefont {Tew}}, \bibinfo {author} {\bibfnamefont {C.}~\bibnamefont {van
  W{\"u}llen}}, \bibinfo {author} {\bibfnamefont {V.~K.}\ \bibnamefont
  {Voora}}, \bibinfo {author} {\bibfnamefont {F.}~\bibnamefont {Weigend}},
  \bibinfo {author} {\bibfnamefont {A.}~\bibnamefont {Wody{\'n}ski}}, \ and\
  \bibinfo {author} {\bibfnamefont {J.~M.}\ \bibnamefont {Yu}},\ }\href
  {\doibase 10.1063/5.0004635} {\bibfield  {journal} {\bibinfo  {journal} {J.
  Chem. Phys.}\ }\textbf {\bibinfo {volume} {152}},\ \bibinfo {pages} {184107}
  (\bibinfo {year} {2020})}\BibitemShut {NoStop}%
\bibitem [{\citenamefont {Gonze}\ and\ \citenamefont {Vigneron}(1989)}]{GV89}%
  \BibitemOpen
  \bibfield  {author} {\bibinfo {author} {\bibfnamefont {X.}~\bibnamefont
  {Gonze}}\ and\ \bibinfo {author} {\bibfnamefont {J.-P.}\ \bibnamefont
  {Vigneron}},\ }\href {\doibase 10.1103/PhysRevB.39.13120} {\bibfield
  {journal} {\bibinfo  {journal} {Phys. Rev. B}\ }\textbf {\bibinfo {volume}
  {39}},\ \bibinfo {pages} {13120} (\bibinfo {year} {1989})}\BibitemShut
  {NoStop}%
\bibitem [{\citenamefont {Van~Gisbergen}\ \emph {et~al.}(1997)\citenamefont
  {Van~Gisbergen}, \citenamefont {Snijders},\ and\ \citenamefont
  {Baerends}}]{GSB97}%
  \BibitemOpen
  \bibfield  {author} {\bibinfo {author} {\bibfnamefont {S.}~\bibnamefont
  {Van~Gisbergen}}, \bibinfo {author} {\bibfnamefont {J.}~\bibnamefont
  {Snijders}}, \ and\ \bibinfo {author} {\bibfnamefont {E.}~\bibnamefont
  {Baerends}},\ }\href@noop {} {\bibfield  {journal} {\bibinfo  {journal}
  {Physical review letters}\ }\textbf {\bibinfo {volume} {78}},\ \bibinfo
  {pages} {3097} (\bibinfo {year} {1997})}\BibitemShut {NoStop}%
\bibitem [{\citenamefont {Zhu}\ \emph {et~al.}(2021)\citenamefont {Zhu},
  \citenamefont {Wang}, \citenamefont {Wang}, \citenamefont {Feng},\ and\
  \citenamefont {Sheng}}]{ZWWFS21}%
  \BibitemOpen
  \bibfield  {author} {\bibinfo {author} {\bibfnamefont {H.}~\bibnamefont
  {Zhu}}, \bibinfo {author} {\bibfnamefont {J.}~\bibnamefont {Wang}}, \bibinfo
  {author} {\bibfnamefont {F.}~\bibnamefont {Wang}}, \bibinfo {author}
  {\bibfnamefont {E.}~\bibnamefont {Feng}}, \ and\ \bibinfo {author}
  {\bibfnamefont {X.}~\bibnamefont {Sheng}},\ }\href {\doibase
  https://doi.org/10.1016/j.cplett.2021.139150} {\bibfield  {journal} {\bibinfo
   {journal} {Chemical Physics Letters}\ }\textbf {\bibinfo {volume} {785}},\
  \bibinfo {pages} {139150} (\bibinfo {year} {2021})}\BibitemShut {NoStop}%
\bibitem [{\citenamefont {Norman}\ \emph {et~al.}(2005)\citenamefont {Norman},
  \citenamefont {Bishop}, \citenamefont {Jensen},\ and\ \citenamefont
  {Oddershede}}]{NBJO05}%
  \BibitemOpen
  \bibfield  {author} {\bibinfo {author} {\bibfnamefont {P.}~\bibnamefont
  {Norman}}, \bibinfo {author} {\bibfnamefont {D.~M.}\ \bibnamefont {Bishop}},
  \bibinfo {author} {\bibfnamefont {H.~J.~A.}\ \bibnamefont {Jensen}}, \ and\
  \bibinfo {author} {\bibfnamefont {J.}~\bibnamefont {Oddershede}},\
  }\href@noop {} {\bibfield  {journal} {\bibinfo  {journal} {The Journal of
  chemical physics}\ }\textbf {\bibinfo {volume} {123}},\ \bibinfo {pages}
  {194103} (\bibinfo {year} {2005})}\BibitemShut {NoStop}%
\bibitem [{\citenamefont {Kjaegaard}\ \emph {et~al.}(2008)\citenamefont
  {Kjaegaard}, \citenamefont {Jorgensen}, \citenamefont {Olsen}, \citenamefont
  {Coriani},\ and\ \citenamefont {Helgaker}}]{KJOCH08}%
  \BibitemOpen
  \bibfield  {author} {\bibinfo {author} {\bibfnamefont {T.}~\bibnamefont
  {Kjaegaard}}, \bibinfo {author} {\bibfnamefont {P.}~\bibnamefont
  {Jorgensen}}, \bibinfo {author} {\bibfnamefont {J.}~\bibnamefont {Olsen}},
  \bibinfo {author} {\bibfnamefont {S.}~\bibnamefont {Coriani}}, \ and\
  \bibinfo {author} {\bibfnamefont {T.}~\bibnamefont {Helgaker}},\ }\href
  {\doibase 10.1063/1.2961039} {\bibfield  {journal} {\bibinfo  {journal} {The
  Journal of Chemical Physics}\ }\textbf {\bibinfo {volume} {129}},\ \bibinfo
  {pages} {054106} (\bibinfo {year} {2008})},\ \Eprint
  {http://arxiv.org/abs/https://doi.org/10.1063/1.2961039}
  {https://doi.org/10.1063/1.2961039} \BibitemShut {NoStop}%
\bibitem [{\citenamefont {Zahariev}\ and\ \citenamefont {Gordon}(2014)}]{ZG14}%
  \BibitemOpen
  \bibfield  {author} {\bibinfo {author} {\bibfnamefont {F.}~\bibnamefont
  {Zahariev}}\ and\ \bibinfo {author} {\bibfnamefont {M.~S.}\ \bibnamefont
  {Gordon}},\ }\href {\doibase 10.1063/1.4867271} {\bibfield  {journal}
  {\bibinfo  {journal} {The Journal of Chemical Physics}\ }\textbf {\bibinfo
  {volume} {140}},\ \bibinfo {pages} {18A523} (\bibinfo {year}
  {2014})}\BibitemShut {NoStop}%
\bibitem [{\citenamefont {Hu}\ \emph {et~al.}(2016)\citenamefont {Hu},
  \citenamefont {Autschbach},\ and\ \citenamefont {Jensen}}]{HAJ16}%
  \BibitemOpen
  \bibfield  {author} {\bibinfo {author} {\bibfnamefont {Z.}~\bibnamefont
  {Hu}}, \bibinfo {author} {\bibfnamefont {J.}~\bibnamefont {Autschbach}}, \
  and\ \bibinfo {author} {\bibfnamefont {L.}~\bibnamefont {Jensen}},\
  }\href@noop {} {\bibfield  {journal} {\bibinfo  {journal} {Journal of
  Chemical Theory and Computation}\ }\textbf {\bibinfo {volume} {12}},\
  \bibinfo {pages} {1294} (\bibinfo {year} {2016})}\BibitemShut {NoStop}%
\bibitem [{\citenamefont {Li}\ \emph {et~al.}(2014)\citenamefont {Li},
  \citenamefont {Suo},\ and\ \citenamefont {Liu}}]{LSL14}%
  \BibitemOpen
  \bibfield  {author} {\bibinfo {author} {\bibfnamefont {Z.}~\bibnamefont
  {Li}}, \bibinfo {author} {\bibfnamefont {B.}~\bibnamefont {Suo}}, \ and\
  \bibinfo {author} {\bibfnamefont {W.}~\bibnamefont {Liu}},\ }\href@noop {}
  {\bibfield  {journal} {\bibinfo  {journal} {The Journal of chemical physics}\
  }\textbf {\bibinfo {volume} {141}},\ \bibinfo {pages} {244105} (\bibinfo
  {year} {2014})}\BibitemShut {NoStop}%
\bibitem [{\citenamefont {Grabo}\ \emph {et~al.}(2000)\citenamefont {Grabo},
  \citenamefont {Petersilka},\ and\ \citenamefont {Gross}}]{GPG00}%
  \BibitemOpen
  \bibfield  {author} {\bibinfo {author} {\bibfnamefont {T.}~\bibnamefont
  {Grabo}}, \bibinfo {author} {\bibfnamefont {M.}~\bibnamefont {Petersilka}}, \
  and\ \bibinfo {author} {\bibfnamefont {E.}~\bibnamefont {Gross}},\
  }\href@noop {} {\bibfield  {journal} {\bibinfo  {journal} {Journal of
  Molecular Structure: THEOCHEM}\ }\textbf {\bibinfo {volume} {501}},\ \bibinfo
  {pages} {353} (\bibinfo {year} {2000})}\BibitemShut {NoStop}%
\bibitem [{\citenamefont {Appel}\ \emph {et~al.}(2003)\citenamefont {Appel},
  \citenamefont {Gross},\ and\ \citenamefont {Burke}}]{AGB03}%
  \BibitemOpen
  \bibfield  {author} {\bibinfo {author} {\bibfnamefont {H.}~\bibnamefont
  {Appel}}, \bibinfo {author} {\bibfnamefont {E.~K.}\ \bibnamefont {Gross}}, \
  and\ \bibinfo {author} {\bibfnamefont {K.}~\bibnamefont {Burke}},\
  }\href@noop {} {\bibfield  {journal} {\bibinfo  {journal} {Physical review
  letters}\ }\textbf {\bibinfo {volume} {90}},\ \bibinfo {pages} {043005}
  (\bibinfo {year} {2003})}\BibitemShut {NoStop}%
\bibitem [{\citenamefont {Elliott}\ \emph {et~al.}(2011)\citenamefont
  {Elliott}, \citenamefont {Goldson}, \citenamefont {Canahui},\ and\
  \citenamefont {Maitra}}]{EGCM11}%
  \BibitemOpen
  \bibfield  {author} {\bibinfo {author} {\bibfnamefont {P.}~\bibnamefont
  {Elliott}}, \bibinfo {author} {\bibfnamefont {S.}~\bibnamefont {Goldson}},
  \bibinfo {author} {\bibfnamefont {C.}~\bibnamefont {Canahui}}, \ and\
  \bibinfo {author} {\bibfnamefont {N.~T.}\ \bibnamefont {Maitra}},\
  }\href@noop {} {\bibfield  {journal} {\bibinfo  {journal} {Chem. Phys.}\
  }\textbf {\bibinfo {volume} {391}},\ \bibinfo {pages} {110 } (\bibinfo {year}
  {2011})}\BibitemShut {NoStop}%
\bibitem [{\citenamefont {Tretiak}\ and\ \citenamefont
  {Chernyak}(2003)}]{TC03}%
  \BibitemOpen
  \bibfield  {author} {\bibinfo {author} {\bibfnamefont {S.}~\bibnamefont
  {Tretiak}}\ and\ \bibinfo {author} {\bibfnamefont {V.}~\bibnamefont
  {Chernyak}},\ }\href {\doibase 10.1063/1.1614240} {\bibfield  {journal}
  {\bibinfo  {journal} {The Journal of Chemical Physics}\ }\textbf {\bibinfo
  {volume} {119}},\ \bibinfo {pages} {8809} (\bibinfo {year} {2003})},\ \Eprint
  {http://arxiv.org/abs/http://dx.doi.org/10.1063/1.1614240}
  {http://dx.doi.org/10.1063/1.1614240} \BibitemShut {NoStop}%
\bibitem [{\citenamefont {Bannwarth}\ and\ \citenamefont
  {Grimme}(2014)}]{BG14}%
  \BibitemOpen
  \bibfield  {author} {\bibinfo {author} {\bibfnamefont {C.}~\bibnamefont
  {Bannwarth}}\ and\ \bibinfo {author} {\bibfnamefont {S.}~\bibnamefont
  {Grimme}},\ }\href {\doibase 10.1016/j.comptc.2014.02.023} {\bibfield
  {journal} {\bibinfo  {journal} {Computational and Theoretical Chemistry}\
  }\textbf {\bibinfo {volume} {1040-1041}},\ \bibinfo {pages} {45} (\bibinfo
  {year} {2014})}\BibitemShut {NoStop}%
\bibitem [{\citenamefont {Ou}\ \emph {et~al.}(2015{\natexlab{b}})\citenamefont
  {Ou}, \citenamefont {Alguire},\ and\ \citenamefont {Subotnik}}]{OAS15}%
  \BibitemOpen
  \bibfield  {author} {\bibinfo {author} {\bibfnamefont {Q.}~\bibnamefont
  {Ou}}, \bibinfo {author} {\bibfnamefont {E.~C.}\ \bibnamefont {Alguire}}, \
  and\ \bibinfo {author} {\bibfnamefont {J.~E.}\ \bibnamefont {Subotnik}},\
  }\href {\doibase 10.1021/jp5057682} {\bibfield  {journal} {\bibinfo
  {journal} {The Journal of Physical Chemistry B}\ }\textbf {\bibinfo {volume}
  {119}},\ \bibinfo {pages} {7150} (\bibinfo {year}
  {2015}{\natexlab{b}})}\BibitemShut {NoStop}%
\bibitem [{\citenamefont {Alguire}\ \emph {et~al.}(2015)\citenamefont
  {Alguire}, \citenamefont {Ou},\ and\ \citenamefont {Subotnik}}]{AOS15}%
  \BibitemOpen
  \bibfield  {author} {\bibinfo {author} {\bibfnamefont {E.~C.}\ \bibnamefont
  {Alguire}}, \bibinfo {author} {\bibfnamefont {Q.}~\bibnamefont {Ou}}, \ and\
  \bibinfo {author} {\bibfnamefont {J.~E.}\ \bibnamefont {Subotnik}},\
  }\href@noop {} {\bibfield  {journal} {\bibinfo  {journal} {The Journal of
  Physical Chemistry B}\ }\textbf {\bibinfo {volume} {119}},\ \bibinfo {pages}
  {7140} (\bibinfo {year} {2015})}\BibitemShut {NoStop}%
\bibitem [{\citenamefont {Adamo}\ and\ \citenamefont {Barone}(1999)}]{PBE0}%
  \BibitemOpen
  \bibfield  {author} {\bibinfo {author} {\bibfnamefont {C.}~\bibnamefont
  {Adamo}}\ and\ \bibinfo {author} {\bibfnamefont {V.}~\bibnamefont {Barone}},\
  }\href {\doibase 10.1063/1.478522} {\bibfield  {journal} {\bibinfo  {journal}
  {J. Chem. Phys.}\ }\textbf {\bibinfo {volume} {110}},\ \bibinfo {pages}
  {6158} (\bibinfo {year} {1999})}\BibitemShut {NoStop}%
\bibitem [{\citenamefont {Weigend}\ and\ \citenamefont {Ahlrichs}(2005)}]{SVP}%
  \BibitemOpen
  \bibfield  {author} {\bibinfo {author} {\bibfnamefont {F.}~\bibnamefont
  {Weigend}}\ and\ \bibinfo {author} {\bibfnamefont {R.}~\bibnamefont
  {Ahlrichs}},\ }\href {\doibase 10.1039/b508541a} {\bibfield  {journal}
  {\bibinfo  {journal} {Physical Chemistry Chemical Physics}\ }\textbf
  {\bibinfo {volume} {7}},\ \bibinfo {pages} {3297} (\bibinfo {year}
  {2005})}\BibitemShut {NoStop}%
\bibitem [{\citenamefont {Mazur}\ and\ \citenamefont
  {W{\l}odarczyk}(2009)}]{MW09}%
  \BibitemOpen
  \bibfield  {author} {\bibinfo {author} {\bibfnamefont {G.}~\bibnamefont
  {Mazur}}\ and\ \bibinfo {author} {\bibfnamefont {R.}~\bibnamefont
  {W{\l}odarczyk}},\ }\href@noop {} {\bibfield  {journal} {\bibinfo  {journal}
  {J. Comput. Chem.}\ }\textbf {\bibinfo {volume} {30}},\ \bibinfo {pages}
  {811} (\bibinfo {year} {2009})}\BibitemShut {NoStop}%
\bibitem [{\citenamefont {Carrascal}\ \emph {et~al.}(2018)\citenamefont
  {Carrascal}, \citenamefont {Ferrer}, \citenamefont {Maitra},\ and\
  \citenamefont {Burke}}]{CFMB18}%
  \BibitemOpen
  \bibfield  {author} {\bibinfo {author} {\bibfnamefont {D.~J.}\ \bibnamefont
  {Carrascal}}, \bibinfo {author} {\bibfnamefont {J.}~\bibnamefont {Ferrer}},
  \bibinfo {author} {\bibfnamefont {N.}~\bibnamefont {Maitra}}, \ and\ \bibinfo
  {author} {\bibfnamefont {K.}~\bibnamefont {Burke}},\ }\href@noop {}
  {\bibfield  {journal} {\bibinfo  {journal} {The European Physical Journal B}\
  }\textbf {\bibinfo {volume} {91}},\ \bibinfo {pages} {142} (\bibinfo {year}
  {2018})}\BibitemShut {NoStop}%
\bibitem [{\citenamefont {Cocchi}\ \emph {et~al.}(2014)\citenamefont {Cocchi},
  \citenamefont {Prezzi}, \citenamefont {Ruini}, \citenamefont {Molinari},\
  and\ \citenamefont {Rozzi}}]{CPRMR14}%
  \BibitemOpen
  \bibfield  {author} {\bibinfo {author} {\bibfnamefont {C.}~\bibnamefont
  {Cocchi}}, \bibinfo {author} {\bibfnamefont {D.}~\bibnamefont {Prezzi}},
  \bibinfo {author} {\bibfnamefont {A.}~\bibnamefont {Ruini}}, \bibinfo
  {author} {\bibfnamefont {E.}~\bibnamefont {Molinari}}, \ and\ \bibinfo
  {author} {\bibfnamefont {C.~A.}\ \bibnamefont {Rozzi}},\ }\href {\doibase
  10.1103/PhysRevLett.112.198303} {\bibfield  {journal} {\bibinfo  {journal}
  {Phys. Rev. Lett.}\ }\textbf {\bibinfo {volume} {112}},\ \bibinfo {pages}
  {198303} (\bibinfo {year} {2014})}\BibitemShut {NoStop}%
\bibitem [{\citenamefont {Fuks}\ \emph {et~al.}(2015)\citenamefont {Fuks},
  \citenamefont {Luo}, \citenamefont {Sandoval},\ and\ \citenamefont
  {Maitra}}]{FLSM15}%
  \BibitemOpen
  \bibfield  {author} {\bibinfo {author} {\bibfnamefont {J.~I.}\ \bibnamefont
  {Fuks}}, \bibinfo {author} {\bibfnamefont {K.}~\bibnamefont {Luo}}, \bibinfo
  {author} {\bibfnamefont {E.~D.}\ \bibnamefont {Sandoval}}, \ and\ \bibinfo
  {author} {\bibfnamefont {N.~T.}\ \bibnamefont {Maitra}},\ }\href@noop {}
  {\bibfield  {journal} {\bibinfo  {journal} {Phys. Rev. Lett.}\ }\textbf
  {\bibinfo {volume} {114}},\ \bibinfo {pages} {183002} (\bibinfo {year}
  {2015})}\BibitemShut {NoStop}%
\bibitem [{\citenamefont {Luo}\ \emph {et~al.}(2016)\citenamefont {Luo},
  \citenamefont {Fuks},\ and\ \citenamefont {Maitra}}]{LFM16}%
  \BibitemOpen
  \bibfield  {author} {\bibinfo {author} {\bibfnamefont {K.}~\bibnamefont
  {Luo}}, \bibinfo {author} {\bibfnamefont {J.~I.}\ \bibnamefont {Fuks}}, \
  and\ \bibinfo {author} {\bibfnamefont {N.~T.}\ \bibnamefont {Maitra}},\
  }\href@noop {} {\bibfield  {journal} {\bibinfo  {journal} {The Journal of
  Chemical Physics}\ }\textbf {\bibinfo {volume} {145}},\ \bibinfo {eid}
  {044101} (\bibinfo {year} {2016})}\BibitemShut {NoStop}%
\bibitem [{\citenamefont {Cave}\ \emph {et~al.}(2004)\citenamefont {Cave},
  \citenamefont {Zhang}, \citenamefont {Maitra},\ and\ \citenamefont
  {Burke}}]{CZMB04}%
  \BibitemOpen
  \bibfield  {author} {\bibinfo {author} {\bibfnamefont {R.~J.}\ \bibnamefont
  {Cave}}, \bibinfo {author} {\bibfnamefont {F.}~\bibnamefont {Zhang}},
  \bibinfo {author} {\bibfnamefont {N.~T.}\ \bibnamefont {Maitra}}, \ and\
  \bibinfo {author} {\bibfnamefont {K.}~\bibnamefont {Burke}},\ }\href
  {\doibase https://doi.org/10.1016/j.cplett.2004.03.051} {\bibfield  {journal}
  {\bibinfo  {journal} {Chemical Physics Letters}\ }\textbf {\bibinfo {volume}
  {389}},\ \bibinfo {pages} {39} (\bibinfo {year} {2004})}\BibitemShut
  {NoStop}%
\end{thebibliography}%

\end{document}